\documentclass[aps,prx,twocolumn,secnumarabic,superscriptaddress]{revtex4-1}

\usepackage[utf8]{inputenc}
\usepackage{longtable}
\usepackage{morefloats}
\usepackage[dvips]{graphicx,color}
\usepackage[dvipsnames]{xcolor}
\usepackage{epsfig,graphicx,amsfonts,amsbsy}
\usepackage{graphicx,amsfonts,amsbsy}
\usepackage{amsmath,amsfonts,amsthm,amssymb}
\usepackage{url}
\usepackage{verbatim}
\usepackage[colorlinks=true,allcolors=blue]{hyperref}
\usepackage{array}
\usepackage{multirow}
\usepackage{tabularx}
\usepackage{multirow}
\usepackage{braket} 
\usepackage{dsfont}
\usepackage{bm}
\usepackage{bbm}
\usepackage{natbib}
\usepackage{multibib}
\usepackage{placeins}
\usepackage{afterpage}

\begin{document}
\title{Crossed Andreev reflection in spin-polarized chiral edge states due to the Meissner effect}
\author{Tam\'as Haidekker Galambos}
\affiliation{Department of Physics, University of Basel, Klingelbergstrasse 82, CH-4056 Basel, Switzerland}
\author{Flavio Ronetti}
\affiliation{Department of Physics, University of Basel, Klingelbergstrasse 82, CH-4056 Basel, Switzerland}
\author{Bence Het\'enyi}
\affiliation{Department of Physics, University of Basel, Klingelbergstrasse 82, CH-4056 Basel, Switzerland}
\author{Daniel Loss}
\affiliation{Department of Physics, University of Basel, Klingelbergstrasse 82, CH-4056 Basel, Switzerland}
\author{Jelena Klinovaja}
\affiliation{Department of Physics, University of Basel, Klingelbergstrasse 82, CH-4056 Basel, Switzerland}
\date{\today}


\begin{abstract}
We consider a hybrid quantum Hall-superconductor system, where a superconducting finger with oblique profile is wedged into a two-dimensional electron gas in the presence of a perpendicular magnetic field, as considered by Lee {\it et al.}, Nat.~Phys.~13, 693 (2017). The electron gas is in the quantum Hall regime at filling factor $\nu=1$. Due to the Meissner effect, the perpendicular magnetic field close to the quantum Hall-superconductor boundary is distorted and gives rise to an in-plane component of the magnetic field.  This component enables nonlocal crossed Andreev reflection between the spin-polarized chiral edge states running on opposite sides of the superconducting finger, thus opening a gap in the  spectrum of the edge states without the need of spin-orbit interaction or nontrivial magnetic textures. We compute numerically the transport properties of this setup and show that a negative resistance exists as a consequence of nonlocal Andreev processes. We also obtain numerically the zero-energy local density of states, which systematically shows peaks stable to disorder. The latter result is compatible with the emergence of Majorana bound states.
\end{abstract}


\maketitle


\section{Introduction}
When a superconducting element is inserted between two normally-conducting regions, non-local electron correlations can be induced~\cite{Tinkham2004,Lesovik2001,Choi2000,Recher2001,Recher2002}. In these hybrid junctions, electrons with an energy below the superconducting gap are transmitted through the superconducting region as holes to the other normal contact, thus leaving a Cooper pair (CP) in the superconductor (SC)~\cite{Byers1995,Melin2004}. This mechanism is called crossed Andreev reflection (CAR)~\cite{Andreev1964,Deutscher2000} and, due to its nonlocal character, is extremely appealing for potential applications in quantum communication~\cite{Sato2010,Hofstetter2009,Herrmann2010,Das2012,Schindele2012,Tan2015,Fulop2015}. Several systems have been proposed as possible candidates to observe CAR~\cite{Torres1999,Giazotto2006,Cayssol2008,LeHur2008,Linder2009,Chen2011,Cottet2012,Klinovaja2012CNT,Sadovskyy2012,Rech2012,Reinthaler2013,Klinovaja2014a,Klinovaja2014b,Wang2015,Hou2016,Beiranvand2017,Reeg2017,Zhang2017,Thakurathi2018,Breunig2018,Finocchiaro2018,Hsu2018,Chen2018,Rosdahl2018,Zhang2019,Blasi2019,Blasi2020,Blasi2020b,Blasi2020c,Ronetti2020,Fuchs2021,Manesco2021} 
and some experimental evidence has been reported~\cite{Beckmann2004,Russo2005,CaddenZimansky2009,Hofstetter2009,Herrmann2010,Das2012,Schindele2012,Tan2015,Fulop2015,Wei2010,Hofstetter2011,Schindele2014,Deacon2015,Lee2017,Baba2018,Ueda2019,Vries2019,Gul2021}. 

In this regard, a prominent role is played by hybrid systems where quantum Hall effect (QHE) and SC correlations co-exist~\cite{Chtchelkatchev2007,Khaymovich2010,Stone2011,Komatsu2012,Rickhaus2012,Calado2015,Amet2016,Beconcini2018,Sekera2018,Draelos2018,Seredinski2019,Zhi2019,Indolese2020,Zhao2020,Veyrat2020,Hatefipour2021}. In particular, QHE edge states at filling factor $\nu=1$, proximitized via CAR processes, are especially promising for the field of topological quantum computation, since they are predicted to host Majorana bound states (MBSs)~\cite{Lindner2012,Clarke2014,Prada2020}. Usually, one assumes that a single chiral edge state is fully spin-polarized along the direction of an applied uniform magnetic field. However, the CAR mechanism, as long as we work with $s$-wave superconductors, is only able to open a gap if this condition of spin polarization is broken. In order to achieve this, previous proposals required materials with strong-spin orbit interaction (SOI) either in the 2DEG or in the superconductor~\cite{Lee2017,Finocchiaro2018,Gul2021}.

\begin{figure}[t!]
	\includegraphics[width=\linewidth]{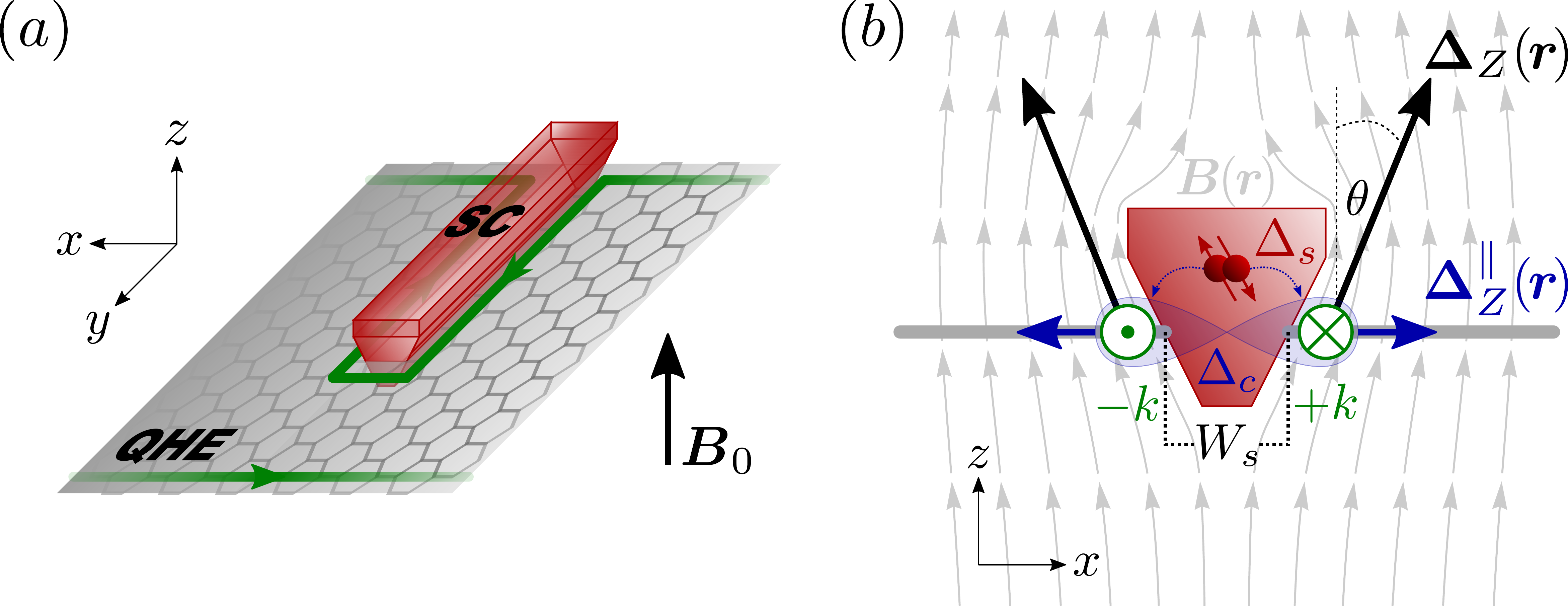}
	\caption{\label{setup}(a) Sketch of the considered setup: a narrow SC finger wedged into a 2DEG tuned into the QHE regime at filling factor $\nu=1$. A single chiral spin-polarized edge state is localized along the system boundaries and around the SC finger.
(b) Side-view schematic of the  oblique cross-section of the SC finger with width $W_s$ at the level of the 2DEG: the locally distorted magnetic field distribution ${\bm{B(\bm r)}}$ due to the Meissner effect. Owing to the oppositely directed in-plane component of the field at the opposite sides of the finger with the counter-propagating chiral (otherwise fully spin-polarized) edge states, the superconducting gap is induced by proximity in the spectrum of the edge states due to  nonlocal CAR processes.}
\end{figure}

In this paper, we show that CAR processes are obtained in spin-polarized edge states as a natural consequence of the geometry of the superconducting element without any need for SOI or nontrivial magnetic textures. We consider a wedge-shaped SC finger inserted into a two-dimensional electron gas (2DEG). We note that the wedge geometry of the superconducting element naturally arises in the fabrication process of hybrid semiconductor-SC or graphene-SC systems where the 2DEG is encapsulated in a different material, e.g., hBN~\cite{Lee2017,Ramezani2021,Gul2021}. A perpendicular magnetic field $B_0$ [see Fig.~\ref{setup}(a)] tunes the 2DEG into the integer quantum Hall regime at filling factor $\nu=1$. Due to the Meissner effect~\cite{Tinkham2004}, magnetic field lines are bent close to the SC, thus effectively generating an in-plane component, as shown in Fig.~\ref{setup}(b). Importantly, the in-plane component has opposite directions at the two opposite QHE-SC interfaces. Hence,  nonlocal CAR processes across the finger generate a proximity-induced superconducting gap in the spectrum of the chiral edge states. We compute numerically this gap, showing that it exhibits an oscillating pattern as a function of the SC region width. As a consequence of CAR, electrons are converted to holes when they are transmitted through the SC and a negative resistance is expected~\cite{Lee2017,Clarke2014}. By means of the recursive Green's function (RGF) method~\cite{MacKinnon1985,Lee1981,Sancho1984,Sancho1985,Meir1992}, we show that in our setup a negative resistance exists due to the presence of CAR processes induced by the Meissner effect without any spin-orbit interaction. In correspondence with these negative values of resistance, we compute the zero-energy local density of states (LDOS). At the open end of the SC finger, we consistently observe a zero-bias peak which is stable with respect to disorder. This result is compatible with the emergence of MBSs~\cite{Clarke2014}. The second MBS merges with the continuum spectrum of propagating QHE edge states and cannot be resolved.


\section{Model}
Inspired by recent experiments~\cite{Lee2017,Gul2021}, we consider a setup consisting of a narrow SC finger wedged into a 2DEG, see  Fig.~\ref{setup}(a).  A strong external magnetic field $B_0$ drives the system into the QHE regime at the filling factor $\nu=1$. The magnetic field configuration is strongly affected by the presence of the SC finger as indicated in Fig.~\ref{setup}(b). We model the system by the effective strictly two-dimensional  Hamiltonian, which describes the 2DEG and a rectangular SC strip of width $W_s$ and length $L_s$,
\begin{equation}
H= \frac{1}{2} \int d^2r  \left[\sum_{i=s,n}\Psi_i^\dagger\mathcal{H}_i\Psi_i +\left(\Psi_s^\dagger\mathcal{T} \Psi_{n}+\mathrm{H.c.}\right)\right],
\label{eq:ModelHamilton}
\end{equation}
given in the Bogoliubov-de Gennes (BdG) formalism, expressed with the basis $\Psi_i=\left[\psi_{i,\uparrow},\psi_{i,\downarrow},\psi_{i,\uparrow}^\dagger,\psi_{i,\downarrow}^\dagger\right]^T$, where $\psi_{i,\sigma}(\bm{r})$ annihilates an electron with spin $\sigma$ at $\bm{r}=\left(x,y\right)$ in the SC ($i=s$) or in the normal ($i=n$) region, respectively.
The Hamiltonian densities $\mathcal{H}_s$ and $\mathcal{H}_{n}$ determine the behavior of the individual subsystems, while $\mathcal{T}$ describes tunnel-coupling between them on the interface along the perimeter of the finger.

In the bulk $s$-wave superconductor, the Hamiltonian density reads
$\mathcal{H}_{s} =\left[-\hbar^2 \bm{\nabla}_{\bm{r}}^2/\left(2m_s\right)-\mu_s\right]\tau_z+\Delta_s\sigma_y\tau_y$, with Pauli-matrices $\sigma_{i}$ and $\tau_{j}$ acting in spin- and Nambu-space, respectively.
The effective mass is given by  $m_s$, the Fermi-level is controlled by the chemical potential $\mu_s$ and the magnitude of the $s$-wave pairing is $\Delta_s$. The SC coherence length is defined as $\xi_s=\hbar^2 k_{F,s} /(m_s\Delta_s)$, where $k_{F,s}=\sqrt{2m_s\mu_s}/\hbar$ is the Fermi momentum of the SC.

\begin{figure}[htb!]
	\includegraphics[width=\linewidth]{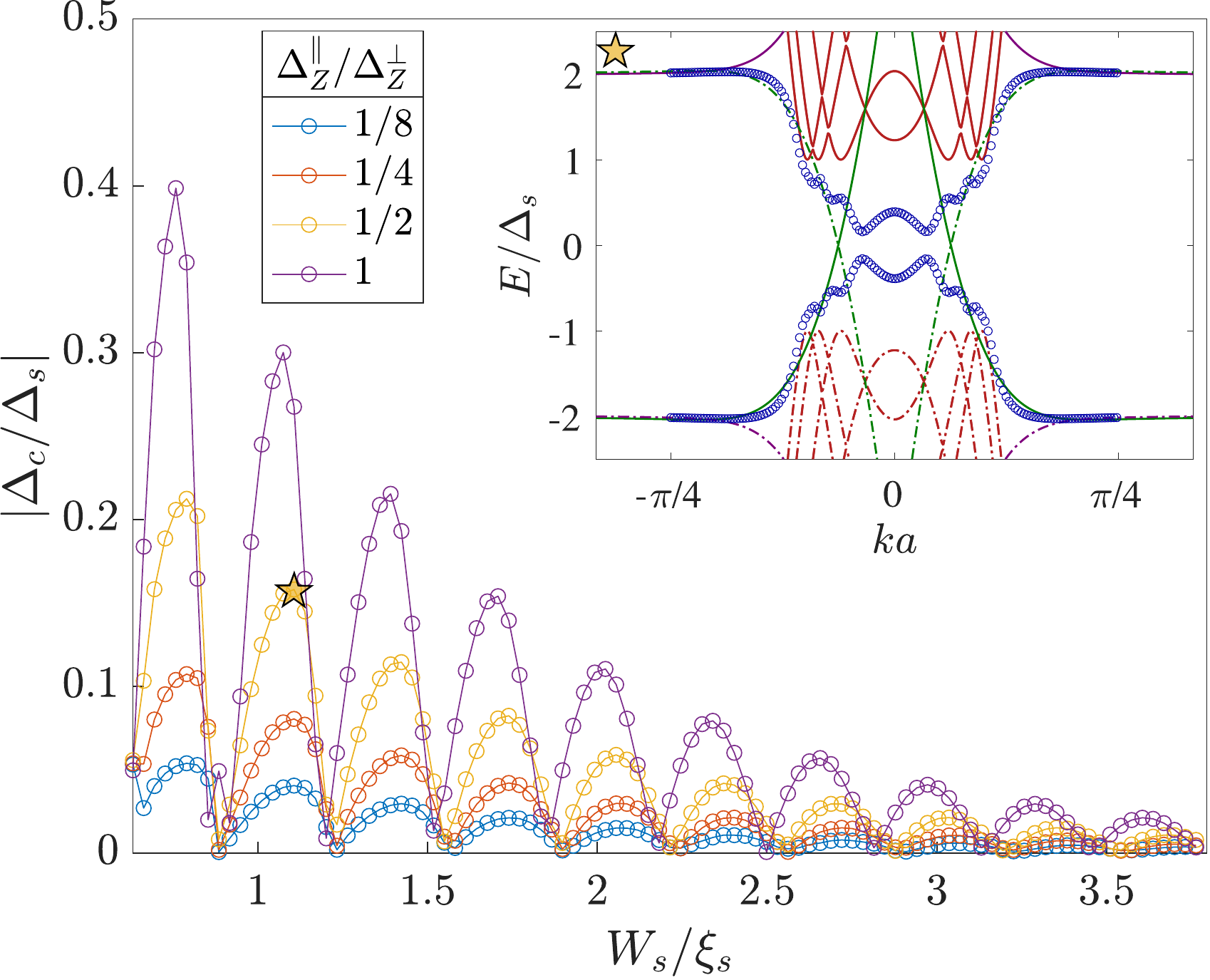}
	\caption{\label{gaposc}
The	proximity-induced CAR gap $\Delta_c$ relative to  SC parent gap $\Delta_s$ as  function of  SC finger width $W_s$ in units of  parent SC coherence length $\xi_s$, for different ratios of  in-plane and out-of-plane components of  Zeeman-energy, $\Delta_Z^\parallel/\Delta_Z^\perp=\left\{1/8,1/4,1/2,1\right\}$. Numerical results are obtained for an infinite QHE-SC-QHE system with $\Delta_Z^\perp=2\Delta_s$, $\mu_s=5\Delta_s$, $\mu_n=3.125\Delta_s$, $m_s=m_n$, a Landau gap in the QHE part of $\Delta_{LL}=\hbar e B_0/m_n=6.28\Delta_s$, along with a perfect interface coupling $\tilde{t}_c$.
(Inset) Low-energy spectrum of the infinite system at $W_s\approx 1.14\xi_s$ and $\Delta_Z^\parallel/\Delta_Z^\perp=1/2$, as indicated by  yellow asterisk in main panel. Red (green) lines correspond to  SC (edge states) in the absence of tunneling. Full lines stand for the electronlike while dash-dotted ones for the holelike parts of the spectra. Blue circles show the spectrum of a coupled QHE-SC-QHE junction, where the edge states are gapped due to CAR. This gap $\Delta_c$ is shown as a function of $W_s$ in the main panel.}
\end{figure}

For the following discussion, a low-energy model, independent of the details of the underlying lattice structure, is sufficient for the description of QHE edge states. In this sense, a quadratic dispersion is appropriate for both semiconductor- and graphene-based systems and the QHE region can be described by $\mathcal{H}_{n}=\mathrm{diag}\left(\tilde{\mathcal{H}}_{n},-\tilde{\mathcal{H}}_{n}^\ast\right)$, with
\begin{equation}
\tilde{\mathcal{H}}_{n}=-\frac{\hbar^2}{2m_{n}}\left[\nabla_{\bm{r}}+\frac{i e}{\hbar}\bm{A}\left(\bm{r}\right)\right]^2-\mu_{n}+\bm{\Delta}_Z\left(\bm{r}\right)\cdot \bm\sigma.
\label{qhamiltonian}
\end{equation}
This 2DEG has effective mass $m_n$ and charge $-e$, with $e>0$ being the unit charge. 
Since only the out-of-plane component contributes to the orbital effect in 2D, the  vector potential $\bm A(\bm r)$ is associated with the $z$-component of the magnetic field, i.e., $B_z=B_0$. The out-of-plane component is homogeneous everywhere in the QHE region and vanishing inside the SC, see supporting results in Appendix~\ref{app:Meissner}. Importantly, the local distortion of the magnetic field induced by the Meissner effect affects the Zeeman term coefficient. The latter is conveniently modeled as a vector with a nonuniform in-plane- and a homogeneous out-of-plane component $\bm\Delta_Z(\bm r)=(\Delta_Z^\parallel\bm d(\bm r)\exp\left(-\left|\bm d(\bm r)\right|/\zeta\right)/\left|\bm d(\bm r)\right|,\Delta_Z^\perp)$.  Here, $\bm d(\bm r)$ is the in-plane distance vector to the closest point of the QHE-SC interface  and $\zeta$ determines the decay of in-plane components away from the finger~\cite{FN2}. 

The chemical potential $\mu_{n}$, along with $B_0$ and $\Delta_Z^\perp$, are tuned such that the system is at $\nu=1$ filling with a single spin-polarized chiral edge mode. Moreover, we note that the spin-splitting of the lowest Landau-level is enhanced by exchange coupling thanks to the strong Coulomb-interaction~\cite{Abanin2006}. Thus we can consider larger effective Zeeman energies than bare $g$-factors would suggest. Finally, the coupling between the subsystems over the shared interface boundary $\partial S$ of the finger is modeled as a tunneling term $\mathcal{T}(\bm r)=\int_{\partial S} d \bm R \,\tilde{t}_c \tau_z\,\delta\left(\bm r-\bm R\right)$ with  
coupling strength $\tilde{t}_c$. 
In order to study the behavior of the composite system, we solve numerically the model by means of a finite difference method (details and used set of parameters in Appendices~\ref{app:Model} and~\ref{app:Parameters}), which yields the numerical results presented in the following. 


\section{Crossed Andreev superconducting gap} Let us first focus on a long and narrow SC finger away from its ends. This is equivalent to working with an infinite lateral QHE-SC-QHE junction, in which the two chiral edge states propagate into opposite directions on the two opposite interfaces with the SC region. For SC widths $W_s$ up to the SC coherence length $\xi_s$, nonlocal CAR processes are available. Earlier proposals necessitated the presence of SOI in the SC area to enable this~\cite{Lee2017,Gul2021,Finocchiaro2018}, whereas in our model, the SC geometry naturally induces an in-plane Zeeman field component via Meissner effect that is oppositely directed on the opposite sides of the SC finger~\cite{FN1}. This CAR process opens the gap $\Delta_c$ in the spectrum of the edge states, see inset of Fig.~\ref{gaposc}. We note that, in principle, another nonlocal process (``cotunneling") could contribute to the gap opening, whereby an electron from one edge tunnels into the other one through the SC. Away from the tip of the superconducting finger, momentum is conserved and cotunneling is always suppressed compared to CAR. This fact is also confirmed in our numerical simulations, in which we do not observe any gap  opening in the spectrum of fully spin-polarized edge states close to the chemical potential.

\begin{figure}[tb!]
	\includegraphics[width=\linewidth]{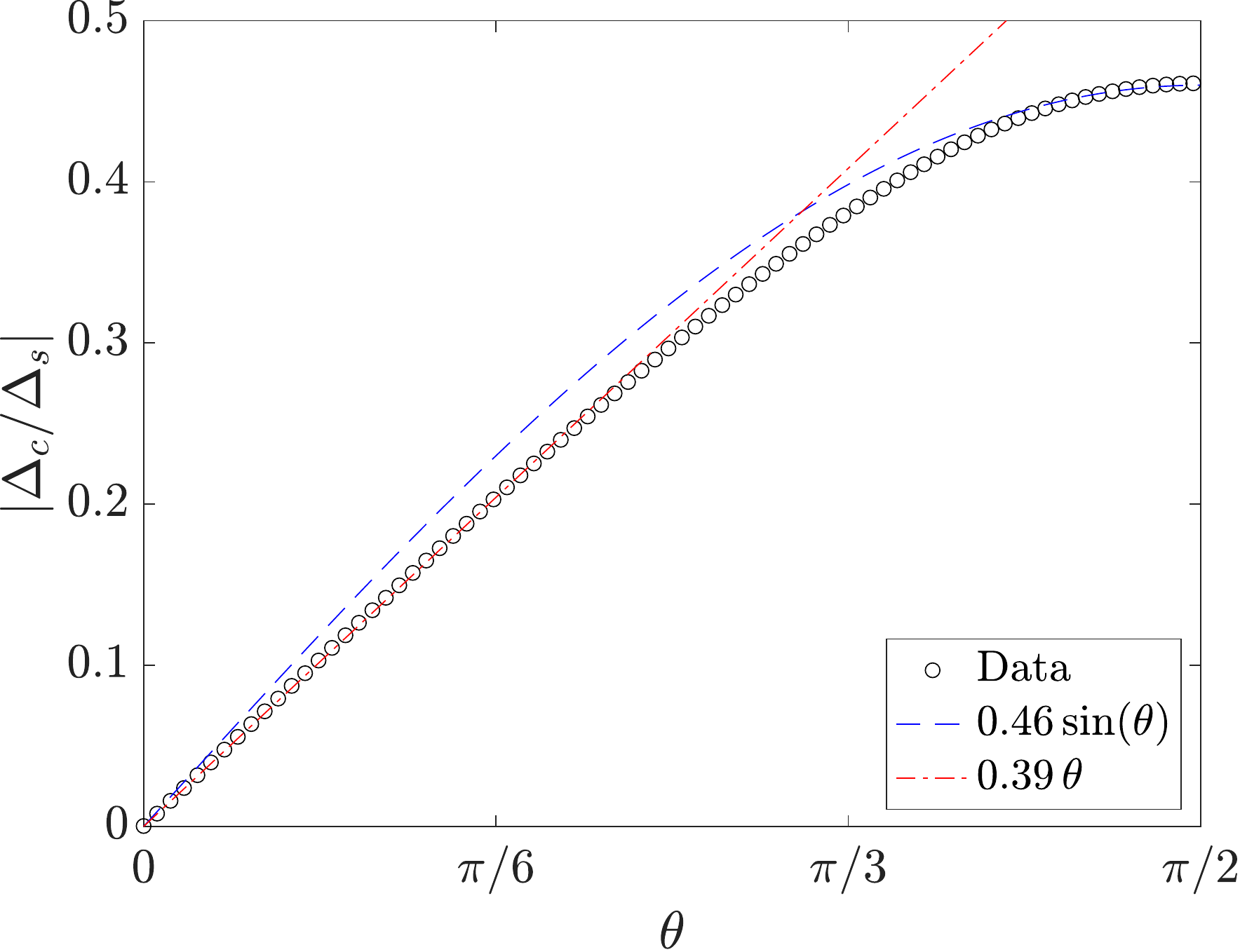}
	\caption{\label{GapInPlaneBSinusMain} The absolute value of the proximity-induced CAR gap $\Delta_c$ relative to SC parent gap $\Delta_s$ as a function of the angle $\theta$, obtained numerically in an infinite QHE-SC-QHE system (see Appendix~\ref{app:Gap} and, in particular,  Fig.~\ref{GapInPlaneBSinus} for further details). The blue and red dotted curves are analytical fits with a sine function and a linear function, respectively, of $\theta$. }
\end{figure}

The main part of Fig.~\ref{gaposc} displays the magnitude of $\Delta_c$ as a function of $W_s$ for different strengths of the in-plane Zeeman component, corresponding to different tilt angles $\theta=\arctan(\Delta_Z^\parallel/\Delta_Z^\perp)$ [see Fig.~\ref{setup}(b)]. Note that, due to the geometrical constraint of the considered setup, the angle is limited to the interval $0<\theta<\pi/2$.  If we neglect additional dependence on the magnitude and inhomogeneity of the Zeeman field, an approximate form of the gap, capturing the essential physics, can be obtained (see Appendix~\ref{app:Gap}):
\begin{equation}
\Delta_c\propto ({\tilde{t}}_c)^2\Delta_s \mathrm\, e^{-W_s/\xi_s}\cos\left(k_{F,s}W_s\right)\sin \theta.\label{eq:gap}
\end{equation}
The dependence on $\theta$ is also easily understood as the spin-overlap of two oppositely canted edge state spins with the spin singlet  of a Cooper pair. In Fig.~\ref{GapInPlaneBSinusMain}, we confirm numerically the validity of this approximate form of the gap (see also Appendix~\ref{app:Gap} for further details) by computing the dependence of $\Delta_c$ on the angle $\theta$. The gap starts linearly with small $\theta$. In addition, the blue dotted line in Fig.~\ref{GapInPlaneBSinusMain}, corresponding to a sine function, fits almost exactly the numerical data, thus justifying the presence of a factor $\sin\theta$ in the induced gap expression. Importantly, $\Delta_c$ is monotonically increasing with $\theta$ in the interval $0<\theta<\pi/2$. We note that in this calculations, we kept the magnitude of the magnetic field constant. The modification of the edge states by the Zeeman term was not taken into account in Eq. (\ref{eq:gap}), which explains small deviations from the predicted behavior. 

\section{Negative resistance}
\begin{figure}[tb!]
\includegraphics[width=\linewidth]{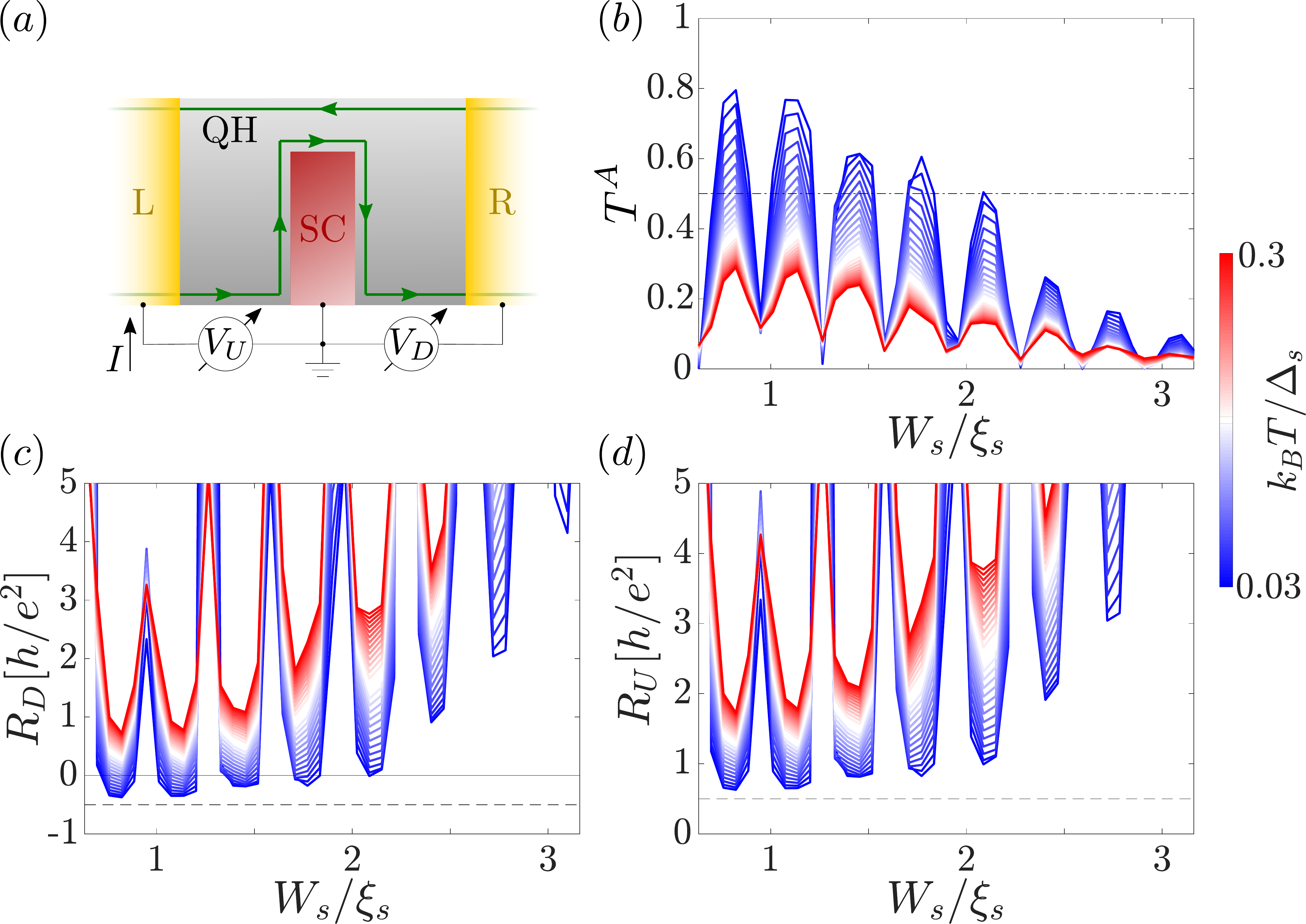}
\caption{\label{transport}(a) Simulated transport measurement arrangement based on those in experiments~\cite{Lee2017,Gul2021,Beconcini2018}. 
(b) Andreev transmission coefficient between the two leads, $T^A$, as a function of the finger width $W_s$ at different temperatures $T$, is used to calculate  (c)  downstream ($R_D$) and (d) upstream ($R_U$) resistances. In both figures, dashed lines indicate the theoretically achievable minimum resistances, $R_{D/U}=\mp R_Q$ for $\nu=1$ filling. 
Results were obtained for the same parameters as the yellow curve in Fig.~\ref{gaposc} and $L_s\approx2.85\xi_s$. For details of numerical simulations, we refer also to Appendix \ref{app:Parameters}.}
\end{figure}

The presence of CAR processes can be tested in the three-terminal transport setup presented in Fig.~\ref{transport}(a)~\cite{Lee2017,Gul2021,Beconcini2018}. Here, a current $I$ is injected through the left lead (L) and drained at the grounded SC, while the right lead (R) is a floating gate. The upstream and downstream potential drops, called respectively $V_U$ and $V_D$, are measured in order to obtain the upstream and downstream resistances, $R_{U/D}=V_{U/D}/I$.~\cite{Amet2016,Lee2017,Gul2021}. 

We calculate numerically the energy-dependent transmission and reflection coefficients by the RGF method~\cite{MacKinnon1985}, by adding left and right leads to our system~\cite{Lee1981,Sancho1984,Sancho1985,Meir1992}.  Due to the chiral nature of the edge states and the insulating bulk of the QHE part, reflection coefficients are zero and an incoming electron is always transmitted from the left to the right. The electron can either undergo a normal transmission process, thus remaining an electron, or a CAR process, thus being converted to a hole. The corresponding normal- and Andreev-transmission probabilities are indicated as $T^N$ and $T^A$, respectively. We plot the latter quantity as a function of the SC width in Fig.~\ref{transport}(b). Note that the CAR transmission exhibits the same oscillating pattern as $\Delta_c$ as a function of $W_s$  in Fig.~\ref{gaposc}. This is a striking evidence of the relation between the gap opening and the presence of nonlocal electron-to-hole converting processes in the system. When temperature is increased, the broadening of Fermi distribution makes the conversion of electrons to holes less effective, thus reducing the CAR transmission probability.

By using the Landauer-B\"uttiker formula for the  three-terminal current-bias relation in the linear-response regime~\cite{Lambert1998,Beconcini2018}, the upstream and downstream resistances can be related to transmission probabilities as $R_{U/D}=R_Q \left(T^N\pm T^A\right)/\left(\nu T^A\right)$, where $R_Q=h/\left(2e^2\right)$ is the quantum of resistance (see Appendix~\ref{app:TransportCalc} for detailed calculations).  In Fig.~\ref{transport}, we present the width dependence of the downstream and upstream resistance as obtained from the transmission probabilities computed numerically. We note that, since normal and Andreev transmission coefficients are related by $T^N+T^A=\nu=1$~\cite{Hou2016,Beconcini2018}, whenever $T^A>0.5$, the low-temperature downstream resistance $R_D$ becomes negative. On the contrary, the upstream resistance is always defined as a positive quantity, satisfying $R_U= R_D+2 R_Q/\nu$.   Interestingly, for smaller widths, at which the CAR processes are more efficient, the upstream and downstream resistances show dips very close to theoretical limits $\pm R_Q$, respectively. From their expressions in terms of transmission probabilities, we can read off that this limit can be exactly achieved only in the case of perfect CAR, i.e., for $T^A=1$~\cite{Zhang2019}.
The negative resistance dips at low temperatures are a direct and clear evidence for the CAR gap induced by the Meissner effect. Alternative explanations for the negative resistance dips, such as the formation of Andreev edge states~\cite{Hoppe2000,Chtchelkatchev2007,Khaymovich2010}, are not applicable for edge states at $\nu=1$. 

As a final comment, we note that negative resistance can be observed in the quantum Hall edge states at $\nu=2$ as well. Nevertheless, for spin-unpolarized edge states at $\nu=2$, we interpret the appearance of negative resistance values as a result of local Andreev reflection processes, as shown in detail in Appendix~\ref{app:TransportResults}. Moreover, at filling factor $\nu=2$ no topological phase is expected, while at $\nu=1$, MBSs are expected to emerge at the two ends of the SC finger~\cite{Clarke2014}.


\section{Majorana bound states} 
\begin{figure}[t!]
\includegraphics[width=\linewidth]{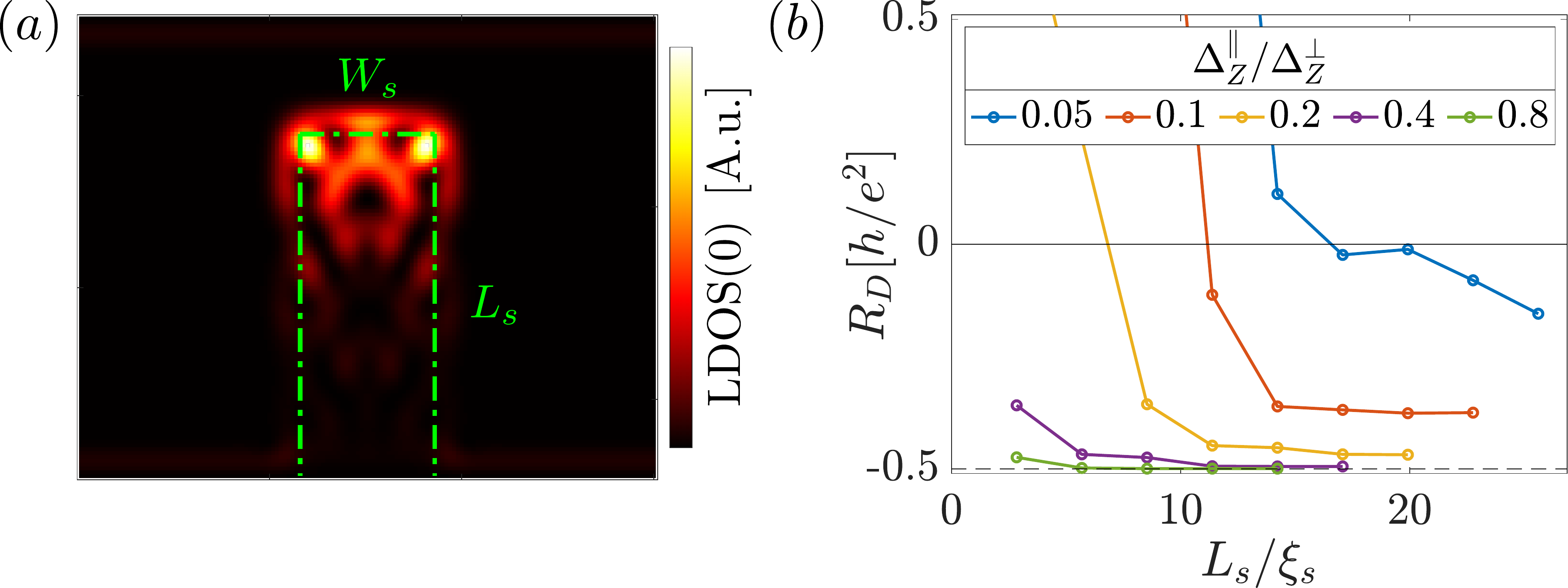}
\caption{\label{ldos}
(a) LDOS at zero energy, LDOS($E=0$), in the finite-geometry finger setup for the same set of parameters as used in Fig.~\ref{transport}. 
Note that edge states along the SC finger gap out and a localized zero-energy MBS forms on the tip of the finger, as well as its delocalized partner in the lower edge states (away from the finger), invisible at the scale adapted to the MBS density. (b) A negative downstream resistance $R_D$ for smaller values of $\Delta_Z^\parallel$ can be achieved by increasing the SC finger length $L_s$. The latter reduces the MBS wavefunction overlap (localization characterized by induced CAR gap $\Delta_c$), and thus also the energy-splitting away from $E=0$, rendering electron-to-hole conversion through the lower MBS (hybridized with the edge state) more efficient  in the transport. All parameters are kept unchanged except for $W_s=0.82\xi_s$ and  $T=0.006\Delta_s$. For details of numerical simulations, we refer also to Appendix \ref{app:Parameters}.}
\end{figure}

The negative peaks in the downstream resistance are a clear evidence that the opposite edge states are proximitized by nonlocal CAR processes. Thus a natural question to ask is whether MBSs arise in our system. Indeed, the presence of two counter-propagating edge states with opposite spin-canting and an induced CAR gap constitute all necessary ingredients to realize a quasi-one-dimensional topological SC~\cite{Kitaev2001,Oreg2010,Lutchyn2010,Klinovaja2012,Klinovaja2013,Clarke2014,Elliott2015,Laubscher2021}. In order to verify the presence of zero-energy MBS peaks in the finite finger geometry, we compute the LDOS at widths $W_s$ corresponding to $R_D<0$, i.e peaks in $\Delta_c$.  We consistently find zero-energy peaks in the LDOS at the tip of the SC finger, as an example in Fig.~\ref{ldos}(a) shows. We tested the stability of these peaks with respect to disorder in the superconducting chemical potential $\mu_s$, finding that they are remarkably stable even for values of disorder comparable with the parent SC gap size, $\Delta_s$ (see Appendix~\ref{app:TransportResults}). This result is compatible with the emergence of MBSs in our setup. The second MBS, expected to appear on the lower end of the finger, cannot be resolved, since it hybridizes with gapless QHE edge states away from the finger. Indeed, when CAR dominates over normal transmission, it is this second MBS on the lower end that resonantly converts incoming electrons into outgoing holes, thus yielding dips with values close to the minimal $R_D\approx-R_Q$~\cite{Clarke2014}. 

Importantly, the emergence of MBSs indicates that our system is in a topological phase. According to Eq.~\eqref{eq:gap}, this topological gap cannot be closed at any finite value of the in-plane magnetic component $\Delta_Z^\parallel$, thus proving that our system stays in the topological phase even in the case of an incomplete Meissner effect in type-II superconductors. Indeed, for finite-size systems,  a negative resistance,  $R_D<0$, can be observed in the simulations even for small $\Delta_Z^\parallel$, provided that the SC finger is long enough to prevent the overlap between MBSs, see Fig.~\ref{ldos}(b). In this regard, we note that the length used in most of our simulations ($L_s\sim 2.85 \xi_s$) is much smaller than realistic values ($L_s\sim 20 \xi_s$ in Ref.~\cite{Lee2017}), except in
Fig.~\ref{ldos}(b) where we were able to simulate the transport up to realistic values $L_s \sim 25 \xi_s$.

Finally, we provide a heuristic derivation of the relation between MBSs and Andreev transmission coefficient. Ideally, for two MBSs exactly at zero energy, the electron-to-hole conversion through one of the MBSs is perfect and Andreev transmission probability becomes $1$. Nevertheless, due to the finite length $L_s$ of the SC finger, the two MBSs always have a finite overlap that is related to the Majorana localization length $l_c$ and which causes the energy of the modes to split away from zero~\cite{Prada2012,Zyuzin2013,Rainis2013,Churchill2013,Albrecht2016}. In this sense, we can provide an approximate expression for the transmission coefficient $T^A$, tuned off-resonance by the Majorana overlap, when $\theta\ll 1$:
\begin{equation}
T^A\sim 1-\exp\left(-\frac{2L_s}{l_c}\right),
\end{equation}
showing that for longer fingers the value of Andreev transmission is increased. As a consequence, for longer $L_s$, the emergence of negative values of downstream resistance is more likely to occur. This is in agreement with Fig.~\ref{ldos}(b). The Majorana localization length can be related to the induced and parent superconducting gaps by $l_c/\xi_s=\Delta_s/\Delta_c$. For small in-plane component of the magnetic field and using the approximate expression for the gap in Eq.~\eqref{eqSM:maggap} for $\theta\approx\Delta_Z^{\parallel} / \Delta_Z^{\perp}\ll 1$, we can estimate the Andreev transmission to be
\begin{equation}
T^A\sim 1-\exp\left(-\frac{2L_s}{\xi_s}\theta\,\mathrm e^{-W_s/\xi_s}\right).
\end{equation}
We define the critical length $L^*_s$ to be the one at which $T_A$ exceeds  $1/2$. Then, from the previous equation, we see that for small in-plane magnetic fields and fixed $W_s/\xi_s$, $L^{*}_s\sim \xi_s/\theta$. This result shows that, for any value of small in-plane component, the critical length $L^{*}_s$ at which $T^A>1/2$ decreases linearly with the in-plane component, in accordance with Fig.~\ref{ldos}(b).


\section{Discussion}
The expulsion of magnetic flux from the superconductor below its critical temperature and critical magnetic field is the well-known Meissner effect~\cite{Meissner1933}. However, most superconductors in practical applications (e.g., NbN) are type II, meaning that a finite magnetic flux can enter the bulk SC through formation of vortices. The vortex formation is driven by the free energy difference between the vortex core energy and the diamagnetic energy of the penetrating superconducting flux quantum $\Phi_0 = \tfrac{h}{2e}$~\cite{Tinkham2004}. Note that the core of the vortex is smaller than the London penetration length $\lambda$ of the magnetic field, i.e., $\xi_s < \sqrt 2 \lambda$. Therefore, if the width $W_s$ of the superconductor is comparable to the diameter of the vortex core ($\sim \xi_s$), the formation of the vortex is energetically not favorable any more. Here, we restricted our discussion to such a scenario, and assume that the flux penetration in the thin SC finger can be modeled by a finite diamagnetic susceptibility $\chi^{}_\text{\tiny{SC}} < 0$, that is still significantly larger than the susceptibility of normal diamagnetic materials.

Furthermore, we note that the ratio of in- and out-of-plane magnetic fields is $B^\parallel / B^\perp \sim 0.05$ even for a susceptibility value as low as $\chi^{}_\text{\tiny{SC}}\sim -0.16$ [see. Fig.~\ref{Meissner2D}(c) in Appendix~\ref{app:Meissner}] confirming that a negative downstream resistance $R_D$ can be observed even if the flux penetration into the SC is relatively large \cite{Lee2017}. 

Finally, we observe that the values of negative resistance obtained in our simulations are of the order of the quantum of resistance $R_Q\sim 13\,$k$\Omega$, while the value measured in experiments is $R_D\sim 50\,\Omega$~\cite{Lee2017}, i.e. roughly 200 times smaller. This implies that possible detrimental effects, such as an incomplete Meissner effect, would not invalidate the qualitative behavior of our prediction.


\section{Conclusions}
We studied a hybrid QHE-SC system formed by a superconductor finger wedged into a two-dimensional electron gas subjected to a strong out-of-plane magnetic field. We focused on the integer QHE regime at filling factor $\nu=1$. This finger geometry was inspired by recent experimental works~\cite{Lee2017,Gul2021}. Due to the geometrical shape of the superconducting element, Meissner effect bends the magnetic field lines close to the superconductor. By simulating this setup with a two-dimensional discrete model, we computed numerically the gap opened by the nonlocal superconducting correlations induced in the two opposite chiral edges by CAR processes. In a three-terminal biased configuration, we numerically evaluated the upstream and downstream resistances and showed that the latter can be negative due to the presence of CAR processes induced by the Meissner effect. In correspondence with these negative values of downstream resistance, the zero-energy local density of states systematically presents disorder-stable peaks, which are compatible with the formation of Majorana bound states. Compared to previous proposals, our setup represents a rather simple implementation of MBSs, since it does not require additional ingredients such as spin-orbit interactions or magnets with non-trivial textures.

The actual experiments on this type of platforms involve the measurement of the downstream resistance to assess the presence of CAR. Nevertheless, by using this type of measurement, one cannot obtain conclusive evidence for the presence of Majorana bound states. Different kinds of measurement schemes can be proposed in this direction in analogy with recent developments in other setups, such as noise~\cite{Liu2015,Liu2015b,Smirnov2019,Smirnov2022} and entropy measurements~\cite{Sela2019,Smirnov2021,Han2022}.

We believe that the presence of the Meissner effect can be assessed directly by means of magnetic scanning techniques e.g. with NV centers~\cite{Huxter2022}. Moreover, one can envisage an experiment tailored to distinguish between the two mechanisms. In this experiment, one realizes several samples where the superconducting wedges have a different angle $\theta$ and measures the downstream resistance. If the observed values of resistance depend on the angle $\theta$, it implies that CAR is significantly affected by the Meissner effect. 


\begin{acknowledgments}
We thank Z. Hou, S. Hoffman, M. Ramezani, and J. Hutchinson for useful discussions.
T.H.G. acknowledges support from the ``Quantum Computing and Quantum Technologies" Ph.D. School of the University of Basel.
This work was supported by the Swiss National Science Foundation and NCCR QSIT.
This project received funding from the European Union’s Horizon 2020 research and innovation program
(ERC Starting Grant, Grant Agreement No. 757725).
\end{acknowledgments}


\appendix


\section{Meissner effect of the SC finger \label{app:Meissner}}

\begin{figure}[tb!]
	\includegraphics[width=\linewidth]{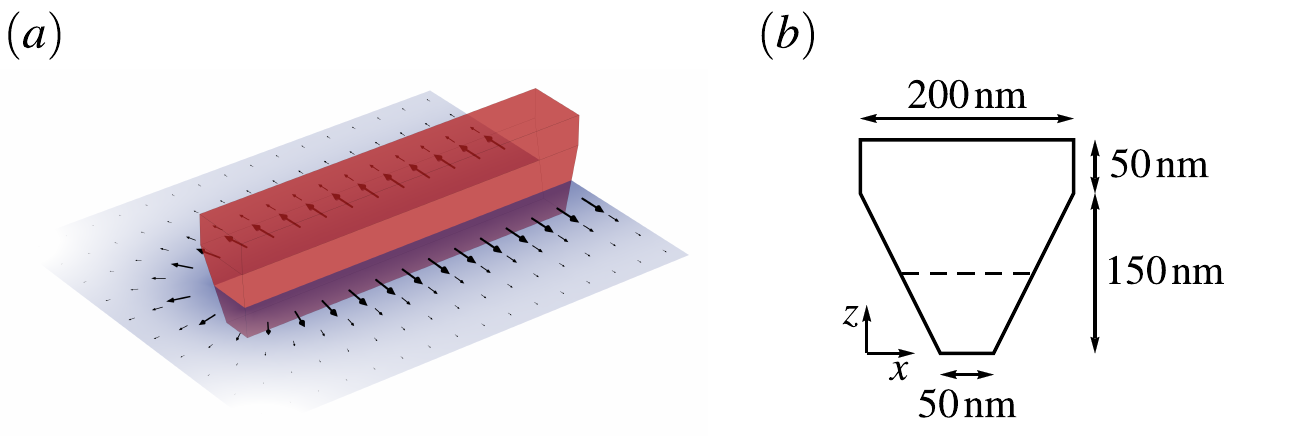}
	\caption{\label{Meissner3D} (a) SC finger (red) and  the in-plane component of the expelled magnetic field due to the Meissner effect. (b) Cross section of the wedge-shaped SC finger and the corresponding dimensions chosen in the simulations. The dashed line, where the width of the SC is $W_s  = 125\,$nm, indicates the plane of the graphene sheet that lies $75\,$nm away from the bottom of the wedge.
	}
\end{figure}

We simulated the Meissner effect using the AC/DC module of COMSOL Multipysics\circledR~\cite{Comsol}. After setting up the geometry as shown in Fig.~\ref{Meissner3D}, we defined the boundary conditions such that the magnetic field $\bf H(\bf r)$ is homogeneous far away from the SC finger, e.g.,
\begin{equation}
{\bf n} \times  {\bf H}({\bf r}) = {\bf n} \times  {\bf H}_0 = 0\,,
\end{equation}
where ${\bf H}_0=H_0 \hat{z}$, with $H_0 = 10^7\,$A/m is the asymptotic magnetic field and $\bf n \parallel \bf{\hat x}$ is the surface normal of the planes at $x = \pm 500\,$ nm, where the boundary condition has been imposed. We note, however, that increasing this value further does not affect the magnetic field close to the SC. Furthermore, periodic boundary conditions were used in the $z$ direction imposed on  surfaces far away.

In the program, we have chosen a ``Physics-controlled mesh" with ``Extremely fine" resolution. The software then solved Maxwell's equations without source terms, e.g.,
\begin{equation}
\nabla \times  {\bf H}({\bf r}) = 0\, ,
\qquad
\nabla \cdot {\bf B}({\bf r}) = 0\, ,
\end{equation}
using a finite element differential equation solver (stationary solver, with relative tolerance $0.01$, and linearity is set to ``Automatic"), where ${\bf B}({\bf r}) = \mu_0 [1+\chi({\bf r})] {\bf H}({\bf r})$ with
\begin{equation}
\chi({\bf r}) = \begin{cases} \chi^{}_\text{\tiny{SC}}\, , &\text{ if } {\bf r} \in D_\text{finger}\, ,\\
0\, , &\text{ if } {\bf r} \not \in  D_\text{finger}\, .
\end{cases}
\end{equation}
Furthermore, we have neglected the magnetic response of h-BN, since its magnetic susceptibility \cite{Crane2000} $\chi^\text{\tiny{hBN}}_\perp = -4.57\times 10^{-4}$ and $ \chi^\text{\tiny{hBN}}_\parallel = -1.27\times 10^{-5}$ is negligible compared to the diamagnetic susceptibility of the SC finger.

We performed the calculation in a 3D model with a finger length of $1.2\,\mu$m and the same wedge angle in the front end as along the sides. The results are shown in Fig.~\ref{Meissner3D}(a). Since the in-plane magnetic field changes only near the end of the finger, we have performed the rest of the simulations in two dimensions along the cross section of the finger with an increased mesh resolution. 

\begin{figure*}[!bth]
	\includegraphics[width=\linewidth]{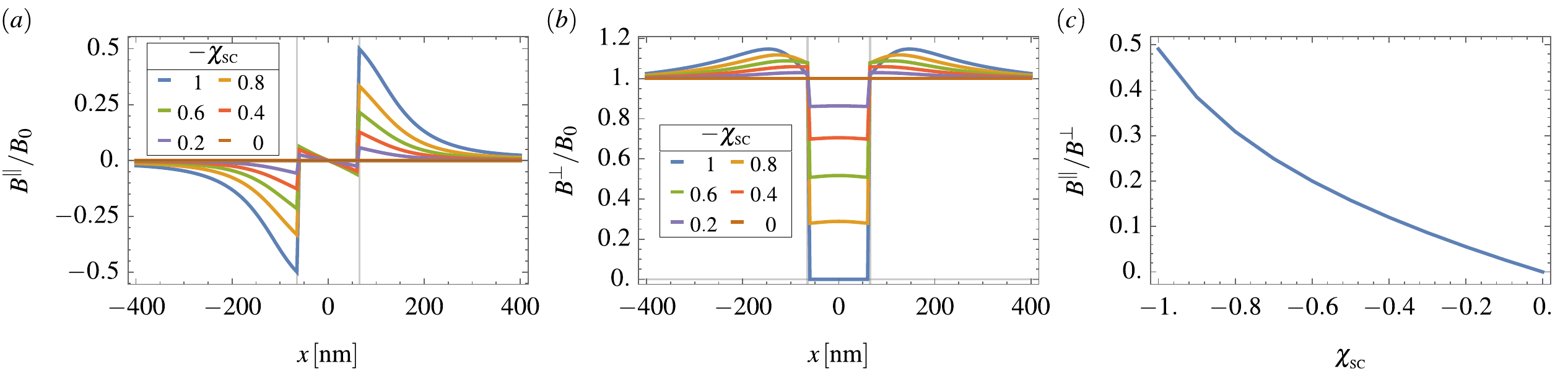}
	\caption{\label{Meissner2D} (a) The in-plane magnetic field $B^\parallel/B_0$ as a function of the $x$-coordinate along the line determined by the dashed line in Fig.~\ref{Meissner3D}(b) for six different magnetic susceptibilities ranging from the perfect diamagnet $\chi^{}_\text{\tiny{SC}}= -1$ to zero magnetic response $\chi^{}_\text{\tiny{SC}} = 0$. The in-plane component $B^\parallel$ is maximum for the perfect diamagnet and is absent if the Meissner effect is neglected. (b) The out-of-plane component of the magnetic field $B^\perp/B_0$ along the same line. The values of $B^\perp$ are only slightly modified close to the interface between the SC and the 2DEG. Thus one can argue that the spatial distribution of the edge state is only slightly affected.  (c) The ratio of the in-plane and out-of-plane components as a function of magnetic susceptibility at $x = 65\,$ nm. For a perfect diamagnet, the ratio could be as large as $B^\parallel/B_0 \approx 1/2$ in this geometry, giving a tilt angle $\theta \approx \pi/4$.}
\end{figure*}

The parallel (in-plane) and perpendicular (out-of-plane) components of the magnetic field as a function of $x$ coordinate are shown in Figs.~\ref{Meissner2D}(a) and~\ref{Meissner2D}(b), where $B_0 = \mu_0 H_0 \sim 12.57\,$T is the strength of the asymptotic out-of-plane $B$ field. We find that the in-plane component decays rapidly and the out-of-plane component is constant outside the SC to a good approximation,
which indicates insignificant out-of-plane flux-focusing, thus edge states in the QHE are not pushed away from the SC and can remain close to the SC-QHE interface. The effect of the in-plane Zeeman field is only relevant close to the SC-QHE interface where the QHE edge state resides. The localization length of edge states is $\ell_\text{B} \sim 7\,$nm. In this region, we model the spatial dependence of the in-plane Zeeman field with an exponential decay towards the bulk QHE system as explained in the main text.


\section{Tight-binding discretization of the QHE-SC hybrid system \label{app:Model}}
In order to study the behavior of the hybrid  QHE-SC system described in the main text, we discretize the continuous Hamiltonian given in Eq.~(\ref{eq:ModelHamilton})  on a square lattice with lattice constant $a$ and solve it  numerically.

In the main text, we only describe the CAR mechanism between $\nu=1$ QHE spin-polarized chiral edges enabled by the presence of in-plane magnetic field components $\Delta_Z^\parallel\neq 0$ as a new, alternative possibility in contrast to the earlier proposed mechanism based on SOI in the SC or the underlying 2DEG substrate.
Here, we add the Rashba-type $\bm{\alpha}\left(\bm k \times\bm \sigma\right)$ SOI with $\bm\alpha\parallel \bm e_z$ (out of the 2D plane) and $|\bm\alpha|=\tilde{\alpha}_s$ into the Hamiltonian density of the SC part, 
$\mathcal H_s\to \mathcal H_s+\mathcal H_{s,\mathrm{SOI}}$, where
\begin{equation}
\mathcal{H}_{s,\mathrm{SOI}}=-i\tilde{\alpha}_s\left(\partial_x\sigma_y-\partial_y\sigma_x\tau_z\right)\tau_z
\end{equation}
in the chosen BdG basis of the main text.
This way we can discuss the discretized models describing either of the possible mechanisms on a common footing and compare their behavior.
In addition, with the same model, we can describe systems with QHE parts at $\nu=2$ filling, which support spin-unpolarized chiral edge states, in which CAR can be induced without the need of any additional mechanism. We will make comparisons between this system and the $\nu=1$ cases on the way. 

The following discretization rules prove to be useful for the derivation of the models presented later on:
\begin{align}
\partial_x\psi(x)&\approx\frac{\psi(x+a)-\psi(x-a)}{2a},\\
\partial_x^2\psi(x)&\approx\frac{\psi(x+a)-2\psi(x)+\psi(x-a)}{a^2},\\
-i\partial_x &\to k_x \to \frac{1}{a}\sin\left(k_x a\right),\\
-\partial_x^2 &\to k_x^2 \to \frac{2}{a^2}\left[1-\cos\left(k_x a\right)\right],
\end{align}
the latter applying to the case of a system with continuous or  discrete (lattice)  translational symmetry, where (crystal)momentum $k_x$ is a good quantum number.
We define the hopping parameters in the QHE and SC regions, $t=\hbar^2/2m_qa^2$ and $t_s=\hbar^2/2m_sa^2$, respectively. Equally, we reinterpret the normal-superconductor coupling $\tilde{t}_c$ of the main text as the coupling hopping $\tilde{t}_c a\equiv t_c$, with energy dimension now. For the same notational convenience, we redefine $\tilde{\alpha}_s/2a\equiv \alpha_s$, also with the dimension of energy. 

Orbital effects in the tight-binding models are incorporated through Peierls-substitution, 
\begin{equation}
t_{ab}\to t_{ab}\exp\left(i\frac{2\pi}{\Phi_0}\int_{\bm{r}_a}^{\bm{r}_b}\bm{A}\cdot d\bm{s}\right),
\end{equation}
where the integration path follows the lattice bonds and $\Phi_0=h/e$ is the regular (nonsuperconducting) flux quantum.


\subsection{Infinite planar QHE-SC-QHE junction geometry}
\begin{figure*}[htb]
	\begin{center}
	\includegraphics[width=\linewidth]{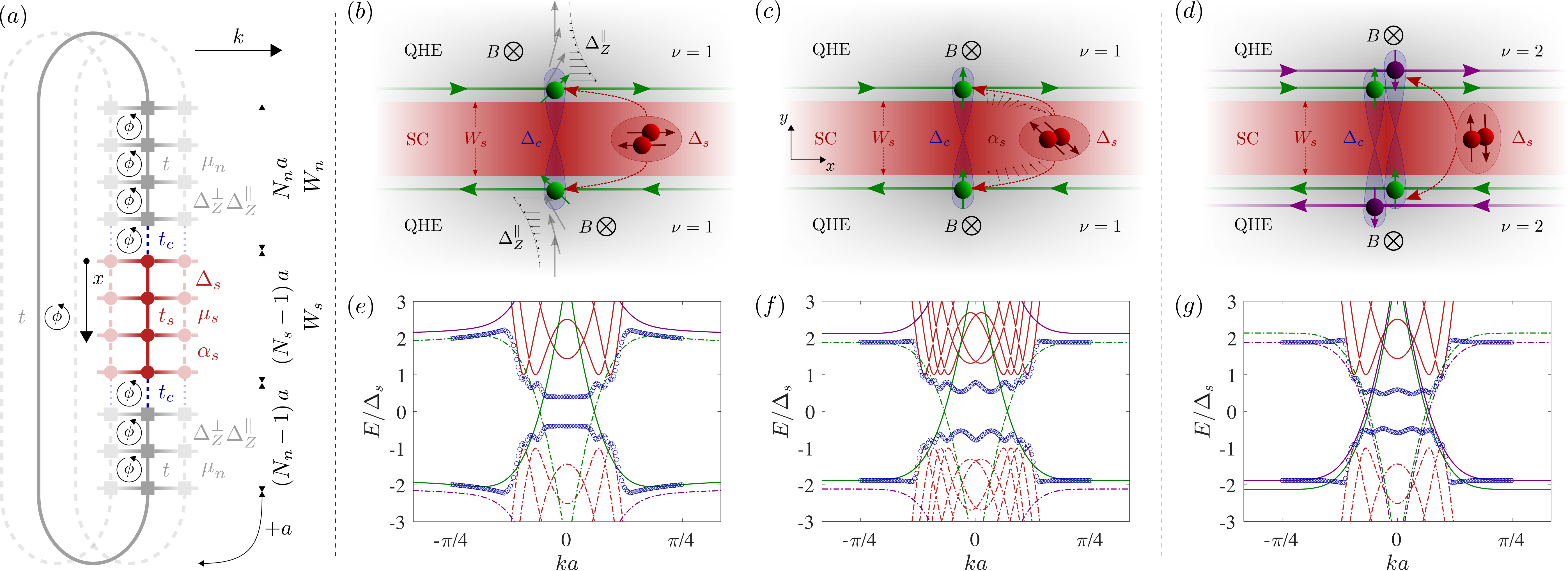}
	\caption{\label{QSQsetup}(a) Quasi-1D discretization scheme for the infinite QHE-SC-QHE structure with periodic boundary conditions in $x$ direction, corresponding to Eqs.~(\ref{eq:qsqs})--(\ref{eq:qsqc}).
		[(b)--(d)] Sketches of the QHE-SC-QHE lateral junction structure with the different possible CAR gap opening mechanisms:
		(b) The case of the main text, where  oppositely directed in-plane magnetic field components in the QHE parts enable the otherwise spin-polarized chiral edges of the QHE at $\nu=1$ filling to be proximitized by the singlet-type SC.
		(c) The mechanism based on SOI considered in earlier works, where the spin-polarized $\nu=1$ edges gain overlap with the singlet structure of CPs by the spin-rotating effect of  SOI in the SC.
		(d) Just for comparison purposes, we show the case of $\nu=2$ filling, where the chiral edges are unpolarized and CAR can take place without any additional ingredient.
		[(e)--(g)] Low-energy spectra with the CAR opened gaps of the respective systems in (b)--(d), obtained in the discretized model of (a).
		Color coding is consistent with the rest of the figure and the main text, green and purple lines show the lowest Landau levels and edge states in the separate QHE subsystem spectrum (with PBC only having the set of chiral edges present next to the SC that are affected by CAR). 
		Red curves represent states in the unconnected SC region with width $W_s$. 
		Full lines mean electronic- and dashed ones hole states.
		Blue circles are the lowest eigenenergies of the hybrid QHE-SC-QHE system at different $k$, showing a significant induced cross-gap $\Delta_c$.
		Parameters used for all (e)--(g): $N_n=50$, $N_s=26$, $t_s=t_c=t$, $\mu_s=0.1t$, $\Delta_s=0.02t$, $\phi=0.01$, and $\zeta=10a$.
		Only in (e), $\mu_n=0.06t$, $\Delta_Z^\perp=0.04t$, $\Delta_Z^\parallel=0.04t$, and $\alpha_s=0$.
		Only in (f), $\mu_n=0.06t$, $\Delta_Z^\perp=0.04t$, $\Delta_Z^\parallel=0$, and $\alpha_s=0.05t$.
		And only in (g), $\mu_n=0.1025t$, $\Delta_Z^\perp=0.0025t$, $\Delta_Z^\parallel=0$,  and $\alpha_s=0$.
	}
	\end{center}
\end{figure*}

To study the energy spectrum and the behavior of CAR induced gap in the spectrum of  chiral edge states coupled to the bulk $s$-wave SC, we first consider the infinite planar QHE-SC-QHE junction geometry (see Fig.~\ref{QSQsetup}), where the system remains translationally invariant in the lateral direction $y$, allowing for the momentum in that direction, $k_y$ (from here onward simply noted as $k$), to be a good quantum number.

We fix our gauge in this geometry as $\bm A(\bm r)=B_0 x\bm e_y$ which is compatible with the translation invariance in $y$ direction and conveniently gives constant $\bm A=0$ in the SC part. This choice is consistent with the assumption that orbital effects are caused by the unchanged out-of-plane magnetic field component $B_0$ in the QHE regions and that $\bm B(\bm r)=0$ in the SC part due to the Meissner effect.

Written in the BdG tight-binding basis of
$C_{l,k}=\left(c_{l,k,\uparrow},c_{l,k,\downarrow},c_{l,-k,\uparrow}^\dagger,c_{l,-k,\downarrow}^\dagger\right)^T$,
where $c_{l,k,\sigma}$ annihilates an electron with $y$-momentum $k$ and spin $\sigma$ on site $x=(l-1)a$, cf. Fig.~\ref{QSQsetup}(a). The tight-binding Hamiltonian of the finite width, infinitely long lateral QHE-SC-QHE system reads $H=H_s+H_n+H_c$, where firstly
\begin{widetext}
\begin{multline}
H_s=\frac{1}{2}\sum_{k}\Bigg(\sum_{l=1}^{N_s}C_{l,k}^\dagger \Big\{\left[4t_s-2t_s\cos\left(k a\right)-\mu_s\right]\tau_z -2\alpha_s\sin\left(k a\right)\sigma_x+\Delta_s\sigma_y\tau_y \Big\}C_{l,k}\\
 +\sum_{l=1}^{N_s-1}\left\{C_{l+1,k}^\dagger \left[-t_s-i \alpha_s\sigma_y\right]\tau_z\ C_{l,k}+\mathrm{H.c.}\right\}\Bigg)
\label{eq:qsqs}
\end{multline}
\end{widetext}
describes the superconducting slab region of width $W_s=(N_s-1)a$ in the middle, with $N_s$ the number of SC sites in the discretized model. Note how the $x$-coordinate origin and the start of numbering of sites is aligned with the upper edge of this region, see Fig.~\ref{QSQsetup}(a).

Secondly, the Hamiltonian of the QHE regions is given as
\begin{widetext}
\begin{multline}
H_n=\frac{1}{2}\sum_{k}\left[\sum_{l=N_s+1}^{N_s+2N_n-1}C_{l,k}^\dagger \Bigg(\left\{4t-2t\cos\left[k a-\tau_z 2\pi\phi\left(l-N_s\right)\right]-\mu_n\right\}\tau_z+ \Delta_Z^\perp\sigma_z\tau_z\right.\\\left. 
+\Delta_Z^\parallel \left\{\exp\left[-\frac{l-\left(N_s+1\right)}{\zeta/a}\right]
-\exp\left[-\frac{\left(N_s+2N_n-1\right)-l}{\zeta/a}\right]\right\}\sigma_x\tau_z\Bigg)C_{l,k} 
-\sum_{l=N_s+1}^{N_s+2N_n-2}\left(C_{l   +1,k}^\dagger\, t\,\tau_z C_{l,k}+\mathrm{H.c.}\right)\right],
\label{eq:qsqq}
\end{multline}
\end{widetext}
where instead of describing two separate QHE slabs on both sides of the SC region, either supporting pairs of chiral states on their inner and outer edges, we model the system wrapped around as shown in Fig.~\ref{QSQsetup}(a), which allows us to get rid of the edge states far from the SC region that would anyway not be affected by the coupling of subsystems.
The described QHE region has $2N_n-1$ sites, but counting the coupling links in Fig.~\ref{QSQsetup}(a) yields $2W_n=2N_n a$ in width (note that magnetic flux is present in the two plaquettes of the coupling region). 
The site dependent part in the cosine is the result of the orbital effect, which with our gauge choice only affects hopping parameters in $y$ direction and $\phi=B_0a^2/\Phi_0$ measures the flux through a plaquette in units of the flux quantum. The extra $\tau_z$ within the cosine takes into account the opposite charge of electrons and holes in the Nambu basis. To respect the integer magnetic unit cell condition with the chosen periodic boundary conditions, $2N_n\phi\in\mathbb{Z}$ has to hold.\\

Finally, the coupling over interfaces between the SC and QHE subsystems reads
\begin{multline}
H_c=\frac{1}{2}\sum_{k}\Big[\Big( -t_c C_{N_s,k}^\dagger\tau_z C_{N_s+1,k}\\  -t_c C_{N_s+2N_n-1,k}^\dagger\tau_z C_{1,k}\Big)+\mathrm{H.c.}\Big].
\label{eq:qsqc}
\end{multline}
We obtain the magnitude of the induced crossed gap $|\Delta_c|$ in the composite system by calculating the lowest positive eigenvalue of the matrix Hamiltonian for a discrete, dense enough set of $k$ values and then minimizing over $k$. This is the lowest point in the upper half of the numerical spectra plotted with blue circles in Figs.~\ref{QSQsetup}(e)--\ref{QSQsetup}(g). We see that it is possible to open a gap in all three considered cases, explained in the sketches of Figs.~\ref{QSQsetup}(b)--\ref{QSQsetup}(d).


\subsection{Finite finger geometry and transport setup}
\begin{figure*}[htb]
	\includegraphics[width=0.75\linewidth]{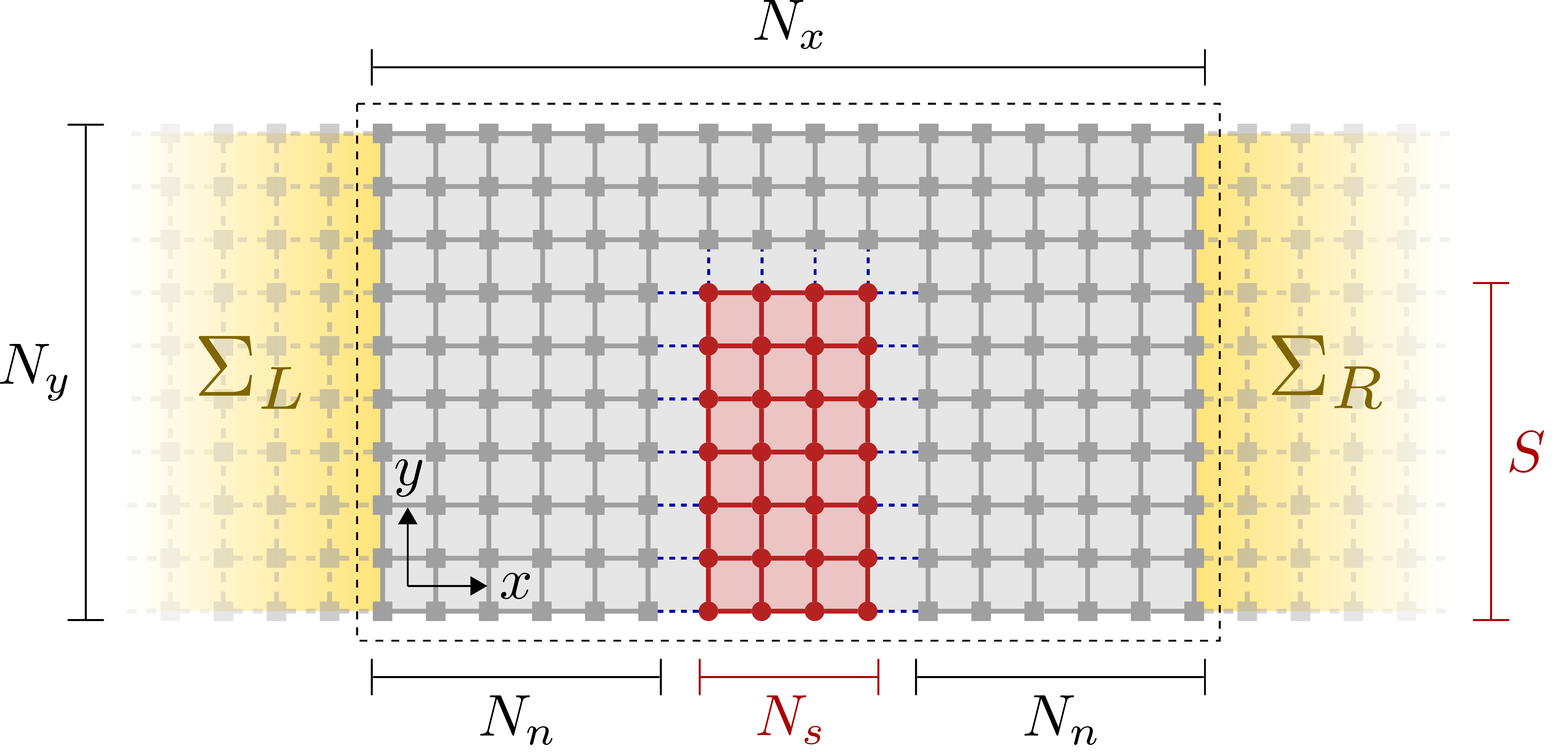}
	\caption{\label{fingersetup}
		Schematic of the discretized finger geometry model with leads indicated that are used for transport calculations. The principal region described by Eq.~(\ref{eq:fingerH}) is marked with the dashed black line and contains $N_x\times N_y$ sites, with $N_x=2N_n+N_s$. The SC finger is $W_s=\left(N_s-1\right)a$ wide and $L_s=\left(S-1\right)a$ long, its sites are marked with red and the red shaded area marks the region of zero magnetic flux through the plaquettes. All gray plaquettes bear a flux, even the region under the blue dashed lines that indicate the interface coupling between SC and QHE, quantified by $t_c$. $\Sigma_{L/R}$ denote the self-energies associated with the left- and right semi-infinite leads, respectively.
	}
\end{figure*}

In order to study the behavior of the finite length SC finger wedged into the QHE insulator, we need its 2D discretized model which we formulate in Eqs.~(\ref{eq:fingerH})--(\ref{eq:fingerD}) based on the scheme shown in Fig.~\ref{fingersetup}.
This time, we choose a different gauge, $\bm A(\bm r)=-B_0 y\bm e_x$, that is compatible with the presence of translationally invariant semi-infinite leads extended along the $x$ direction.
The Hamiltonian
\begin{widetext}
\begin{multline}
H=\sum_{l=1}^{N_x}\sum_{m=1}^{N_y}\sum_{\sigma\sigma'}c_{l,m,\sigma}^\dagger E_{l,m}^{\sigma,\sigma'}c_{l,m,\sigma'}
- \sum_{l=1}^{N_x-1}\sum_{m=1}^{N_y}\sum_{\sigma\sigma'}\left(c_{l+1,m,\sigma}^\dagger t_{l+1,m,l,m}^{\sigma,\sigma'}c_{l,m,\sigma'}+\mathrm{h.c.}\right)\\
- \sum_{l=1}^{N_x}\sum_{m=1}^{N_y-1}\sum_{\sigma\sigma'}\left(c_{l,m+1,\sigma}^\dagger t_{l,m+1,l,m}^{\sigma,\sigma'}c_{l,m,\sigma'}+\mathrm{h.c.}\right)
+\frac{1}{2}\sum_{l=1}^{N_x}\sum_{m=1}^{N_y}\Delta_{l,m}\left(c_{l,m,\uparrow}c_{l,m,\downarrow}-c_{l,m,\downarrow}c_{l,m,\uparrow}+\mathrm{H.c.}\right)
\label{eq:fingerH}
\end{multline}
\end{widetext}
describes the principal region without resorting to the BdG formalism, $c_{l,m,\sigma}$ annihilates an electron on site $(l,m)$ with spin $\sigma$.
The first, onsite term reads
\begin{widetext}
\begin{equation}
E_{l,m}=
\begin{cases}
4t-\mu_n+\Delta_Z^\perp \sigma_z-\Delta_Z^\parallel\mathrm e^{-a\left(N_n-l\right)/\zeta}\sigma_x,
&  1\leq l\leq N_n\  \text{and}\ 1\leq m\leq S,\\
4t-\mu_n+\Delta_Z^\perp \sigma_z  +\Delta_Z^\parallel\,\mathrm e^{-a\left(\sqrt{\left(N_n-l+1\right)^2+(m-S)^2}-1\right)/\zeta} & \\
\times \frac{\left(l-N_n-1\right)\sigma_x+(m-S)\sigma_y}{\sqrt{\left(N_n-l+1\right)^2+(m-S)^2}}, & 1\leq l\leq N_n\  \text{and}\ S< m\leq N_y,\\
4t_s-\mu_s, &  N_n< l\leq N_n+N_s\  \text{and}\ 1\leq m\leq S,\\
4t-\mu_n+\Delta_Z^\perp \sigma_z+\Delta_Z^\parallel\mathrm e^{-a\left(m-S-1\right)/\zeta}\sigma_y, &  N_n< l\leq N_n+N_s\  \text{and}\ S< m\leq N_y,\\
4t-\mu_n+\Delta_Z^\perp \sigma_z+\Delta_Z^\parallel\mathrm e^{-a\left(l-N_n-N_s-1\right)/\zeta}\sigma_x, &  N_x-N_n< l\leq N_x\  \text{and}\ 1\leq m\leq S,\\
4t-\mu_n+\Delta_Z^\perp \sigma_z +\Delta_Z^\parallel\,\mathrm e^{-a\left(\sqrt{\left(l-N_s-N_n\right)^2+(m-S)^2}-1\right)/\zeta}& \\
\times\frac{\left(l-N_s-N_n\right)\sigma_x+(m-S)\sigma_y}{\sqrt{\left(l-N_s-N_n\right)^2+(m-S)^2}}, & N_x-N_n< l\leq N_x\  \text{and}\ S< m\leq N_y,
\end{cases}
\end{equation}
\end{widetext}
while the hopping term in the $x$ direction is given by 
\begin{equation}
t_{l+1,m,n,m}=
\begin{cases}
t_c\mathrm e^{-i2\pi\phi(m-1)}, &   \begin{aligned}l= N_n\  \text{and}\\ 1\leq m\leq S,\end{aligned}\\
t_s-i\alpha_s\sigma_y, &   \begin{aligned}&N_n<l< N_n+N_s\  \text{and}\\ &1\leq m\leq S,\end{aligned}\\
t_c\mathrm e^{-i2\pi\phi(m-1)}, &   \begin{aligned}&l= N_n+N_s\  \text{and}\\ & 1\leq m\leq S,\end{aligned}\\
t\mathrm e^{-i2\pi\phi(m-1)}, &  \text{otherwise}.
\end{cases}
\end{equation}
Similarly, hopping in $y$ direction is described as
\begin{equation}
t_{l,m+1,l,m}=
\begin{cases}
t_s+i\alpha_s\sigma_x, &   \begin{aligned}&N_n<l< N_n+N_s\  \text{and}\\& 1\leq m\leq S,\end{aligned}\\
t_c, &  \begin{aligned}& N_n < l \leq N_n+N_s\  \text{and}\\&  m=S,\end{aligned}\\
t, &  \text{otherwise},
\end{cases}
\end{equation}
and finally, the pairing term has the form
\begin{equation}
\Delta_{l,m}=
\begin{cases}
\Delta_s, &   N_n<l< N_n+N_s\  \text{and}\ 1\leq m\leq S,\\
0, &    \text{otherwise}.
\end{cases}
\label{eq:fingerD}
\end{equation}
Without the leads, the system supports chiral edge states running along the edge of the principal region and around the SC finger, where locally they can gap out due to CAR. If the leads are included, the system becomes infinite and the oppositely moving chiral edges on the top and bottom propagate infinitely without closing in on each other.


\section{Parameter choices and dimensionless  quantities \label{app:Parameters}}

Our intention is far from trying to exactly model experiments~\cite{Lee2017,Gul2021}, but we choose our numerical parameter ranges inspired by them. These setups have a SC finger of length $L_s\sim 1\,\mu$m and width $W_s\sim 50,\ldots, 200\,$nm, thus we assume the region of interest to be $\sim 1\,\mu \mathrm{m}^2$, which we choose to discretize on a square lattice with lattice constant $a\sim 2\,$nm for reasonable computational needs (we eventually considered also smaller model sizes to speed up calculations).

The experiments operate at $B_0\sim 8\ldots 14\,$T magnetic fields: the convenient value of $\phi=a^2B_0/\Phi_0=0.01$, that we use throughout, corresponds to about $B_0\sim 10\,$T. This yields the dimensionless magnetic length in our model with quadratic dispersion $\ell_B/a=\sqrt{\hbar/eB_0}/a=1/\sqrt{2\pi\phi}\sim 4$. In addition, it is reported that the LL gap in the QHE system is $\Delta_{LL}\sim 100\,$meV, which, with its expression $\Delta_{LL}=eB_0/m_q=4\pi\phi \, t$, determines the hopping parameter as $t\sim 0.8\,$eV and we have $\Delta_{LL}/t\sim 0.125$. For the sake of simplicity, we use $t_s=t_c=t$, unless indicated otherwise.

Further, the Fermi velocity in the NbN SC is reported to be $v_{F,s}=1.8\times10^6\,$m/s, which by the definition $\mu_s=m_s v_{F,s}^2/2$ determines the chemical potential in the SC as $\mu_s/t=\left(\hbar v_{F,s}/2at\right)^2\sim 0.13$, we have chosen to use $\mu_s/t=0.1$ in most cases. The following dimensionless expression of the Fermi momentum $k_{F,s}a=\sqrt{\mu_s/t}$ proves useful in the study of the oscillatory behavior of the CAR gap, the Fermi length is $\lambda_{F,s}/a=2\pi/k_{F,s}a\sim 20$ for the $\mu_s/t=0.1$ choice.

The superconducting coherence length in the NbN SC finger of the experiment is estimated to be $\xi_s=(52\pm 2)\,$nm, which, in the simple BCS meanfield approach of our model, reads $\xi_s/a=\hbar v_{F,s}/\Delta_s a = 2\sqrt{\mu_s/t}/\left(\Delta_s/t\right)\sim 26$, meaning a dimensionless gap size $\Delta_s/t\sim 0.024$, instead of which we choose to use the value $\Delta_s/t=0.02$, which corresponds to $\xi_s\approx 31.6a\approx 63\,$nm with  the parameter choices described so far.

The temperature ranges we use in most of the transport calculations, $T/\Delta_s=0.03$--$0.3$, with the critical temperature of the superconducting NbN indicated as $T_c\approx 12\,$K, correspond to a range of approximately $360\,\mathrm{mK}$--$3.6\,\mathrm{K}$. Note that Ref.~\onlinecite{Lee2017} reports measurements in the range of $1.8$--$10\,$K, whereas Ref.~\onlinecite{Gul2021} at $15\,$mK.

For SOI in the SC, we do not have an estimate of the experimental strength, we use the range of values $\alpha_s/t=0.025$--$0.1$ (or $\alpha_s=0$, e.g., throughout the main text), which corresponds to spin-orbit lengths $\lambda_{SO,s}=2\pi /k_{SO,s}\sim 125$--$500\,$nm, where $k_{SO,s}a=\alpha_s/t$ is the dimensionless spin-orbit momentum. If the SOI is present in the SC,  the Fermi momenta are modified as $k_{F,s}^{\pm}a=\sqrt{\mu_s/t+\left(\alpha_s/t\right)^2}\pm \alpha_s/t$. The average Fermi momentum $k_{F,s}^{\pm}a =\sqrt{\mu_s/t+\left(\alpha_s/t\right)^2} $ stays almost unchanged: with our parameters it amounts to less than 5\%.
To form the QHE system at $\nu=1$ filling, we tune the chemical potential to the lowest LL with $\mu_n/t=0.0625$ and introduce the out-of-plane Zeeman-splitting $\Delta_Z^\perp/t=0.04$. This is a rather large value (twice the SC gap $\sim 4\,$meV) if one wanted to achieve it paramagnetically through usual $g$-factor values, however, in case of graphene for example, interaction effects can be very strong, introducing ferromagnetic ordering (still following the external field's direction) that enhance the effective Zeeman strength to hundreds of Kelvins~\cite{Abanin2006}. The experimental fact of observing stable $\nu=1$ QHE plateaus justifies these arguments (along with a similarly large valley splitting in case of graphene). 
For the $\nu=2$ filling, we conveniently choose the same velocities in the edge states, as in the $\nu=1$ case, requiring $\mu_n/t=0.1025$ with simply $\Delta_Z^\perp/t=0$ for the simplest description. For the in-plane component strength, we consider the range $\Delta_Z^\parallel/\Delta_Z^\perp=0\ldots 1$, describing different tilt angles of the oblique wedge walls and/or different Meissner-effect efficiencies.

{\emph{Numerical parameters of figures in the main text.}}

Figure~1.(b). Magnetic field lines are obtained through a COMSOL Multipysics\circledR~\cite{Comsol} simulation (see corresponding section for all parameters), for a wall tilt angle $\tan\left(\theta\right)=1/2$ with perfect Meissner-effect, $\chi=-1$.

Figure~2. $N_n=50$, $N_s=20,\ldots, 120$, $t_s=t_c=t$, $\mu_s=0.1t$, $\Delta_s=0.02t$, $\phi=0.01$, $\mu_n=0.0625t$, $\Delta_Z^\perp=0.04t$, $\Delta_Z^\parallel=\left\{0.04t,0.02t,0.01t,0.05t\right\}$, $\zeta=10a$, and $\alpha_s=0$. The gap is measured by searching the minimum of positive eigenenergies in the range $ka=0\ldots \pi/6$ divided linearly into 500 discretized momentum values.
Inset: Same parameters as the main figure, except for $N_s=36$, $\Delta_Z^\parallel=0.02t$ and the spectrum is displayed in 400 discrete momentum values linearly sampled for the range $-\pi/3\leq ka\leq\pi/3$.

Figure~3.  $N_x=150$, $N_y=120$, $N_s=20,\ldots, 100$, $N_n=\left(N_x-N_s\right)/2$, $S=90$, $t_s=t_c=t$, $\mu_s=0.1t$, $\Delta_s=0.02t$, $\phi=0.01$, $\mu_n=0.0625t$, $\Delta_Z^\perp=0.04t$, $\Delta_Z^\parallel=0.02t$, $\zeta=10a$, and $\alpha_s=0$. For a small linewidth in the RGF calculation, we use $\eta=10^{-5}\Delta_s$, the transmission spectrum is recorded for energies $-\Delta_s\leq E\leq \Delta_s$ with $0.01\Delta_s$ resolution and then resistances are calculated for the range of $k_B T/\Delta_s=0.03,\ldots, 0.3$ with again $0.01\Delta_s$ resolution. 

Figure~4. (a) We use the exact same parameters as in Fig.~3., but the LDOS is calculated only for $E=0$ and $N_s=36$. 
(b)  Downstream resistance values are obtained in systems with parameters $N_x=100$, $N_y=S+30$, $N_s=26$, $N_n=\left(N_x-N_s\right)/2$, $S=\left\{90,180,270,360,450,540,630,720,810\right\}$, $t_s=t_c=t$, $\mu_s=0.1t$, $\Delta_s=0.02t$, $\phi=0.01$, $\mu_n=0.0625t$, $\Delta_Z^\perp=0.04t$, $\Delta_Z^\parallel/\Delta_Z^\perp=\left\{0.05,0.1,0.2,0.4,0.8\right\}$, $\zeta=10a$,  $\alpha_s=0$, $\eta=10^{-6}\Delta_s$, in which transmission spectra are recorded in the range $-0.03\Delta_s\leq E\leq 0.03\Delta_s$ with $\Delta E=0.0001\Delta_s$ granularity. Resistance values are calculated for $k_B T/\Delta_s=0.006$, which, according to the previous section's considerations, corresponds to $T\sim 72\,$mK.


\section{Properties of the induced crossed gap $\Delta_c$ \label{app:Gap}}
In this section, we discuss some additional properties of the gap $\Delta_c$ induced by  crossed Andreev reflection (see Fig. 2 of the main text). We justify the approximate formula for the gap given in Eq. (3) and we use it to provide a physical insight for all the dependencies on system parameters. We also provide some numerical results for the gap dependencies in agreement with the approximate formula. Further, stability of the gap is  tested against parameter variations and disorder effects. Throughout this section, we will compare the case presented in the main text with both the SOI case at $\nu=1$ and the non spin-polarized edge states at $\nu=2$.

We start by presenting the approximate expressions for the induced gap~\cite{Reeg2017} 
\begin{align}
\Delta_{c,\nu=1}^{Z_\parallel}&\propto t_c^2\Delta_s\, f\left(k_{F,s},\xi_s,W_s\right)\sin\left(\theta\right),\label{eqSM:maggap}\\
\Delta_{c,\nu=1}^{\alpha_s}&\propto  t_c^2\Delta_s \, f\left(k_{F,s},\xi_s,W_s\right)\sin\left(k_{SO,s}W_s\right),\label{eqSM:SOIgap}\\
\Delta_{c,\nu=2} &\propto t_c^2\Delta_s \, f\left(k_{F,s},\xi_s,W_s\right)\label{eqSM:nu2},
\end{align}
where
\begin{equation}
f\left(k_{F,s},\xi_s,W_s\right)=\frac{\sinh\left(W_s/\xi_s\right)\cos\left(k_{F,s}W_s\right)}{\cosh\left(2W_s/\xi_s\right)-\cos\left(2k_{F,s}W_s\right)},
\label{eqSM:gapOsc0}
\end{equation}
which in the limit $W_s>\xi_s$ becomes
\begin{equation}
f\left(k_{F,s},\xi_s,W_s\right)\xrightarrow{W_s>\xi_s}\exp\left(-W_s/\xi_s\right)\cos\left(k_{F,s}W_s\right).
\label{eqSM:gapOsc}
\end{equation}
We recall from the main text that $\theta=\arctan(\Delta_Z^\parallel/\Delta_Z^\perp)$, with $0\le\theta<\pi/2$ because of geometrical constraints.

As a first comment, we note that all three crossed gap expressions share the common factor, $t_c^2 \Delta_s \,f\left(k_{F,s},\xi_s,W_s\right)$. The quadratic dependence on $t_c$ is due to the fact that, at least in the perturbative regime $t_c/t<1$, crossed Andreev reflection is a mechanism that involves two tunneling processes between SC and edge states. The linearity in the superconducting parent gap $\Delta_s$ is typical of proximity induced gap and, since all the remaining terms in the expression are smaller than $1$, it sets the maximum value of the induced gap. The negative exponential term in Eq.~\eqref{eqSM:gapOsc}, obtained for the $W_s>\xi_s$ limit, corresponds to the overlap between edge states, separated by a distance $W_s$, mediated by Cooper pairs, whose typical size is set by the SC coherence length $\xi_s$. For narrow fingers, $W_s<\xi_S$, the corresponding full expression, captured by the fraction in the same equation, describes extra resonances as the cosine in the denominator becomes relevant. Finally, the last cosine term induces oscillations with the Fermi wave-vector of the SC due to a definite Fermi-surface and the finite width of the finger: with confinement in the $x$ direction,  
the SC preferably accommodates electronic states with a Fermi wave-length $\lambda_{F,s}=2\pi/k_{F,s}$ commensurate with the width $W_s$ and with reduced probability those states that have incommensurate $\lambda_{F,s}$.
Below, we will show some numerical results related to this oscillatory dependence on $k_{F,s}$.

The remaining sine functions in Eqs.~\eqref{eqSM:maggap} and~\eqref{eqSM:SOIgap} are related to the spin-tilting of the otherwise spin-polarized edge states at $\nu=1$. Since the edge states at $\nu=2$ are not spin-polarized, the crossed gap can be opened without any spin-tilting mechanism and this additional factor is absent in Eq.~\eqref{eqSM:nu2}. In the SOI case at $\nu=1$, the spin-orbit length $l_{SO,s}=2\pi /k_{SO,s}$ determines the typical distance over which a spin flips its direction. Then, the relative spin-tilting on opposite edges separated by the distance $W_s$ is governed by the ratio $W_s/l_{SO,s}$ and determines the opening of the gap through the sine dependence in Eq.~\eqref{eqSM:SOIgap}. In the next subsection, we justify the presence of the sine function term for the in-plane magnetic field case  and we also present numerical simulations to support it. 


\subsection{In-plane magnetic field dependence for $\nu=1$}

\begin{figure}[tb!]
	\includegraphics[width=\linewidth]{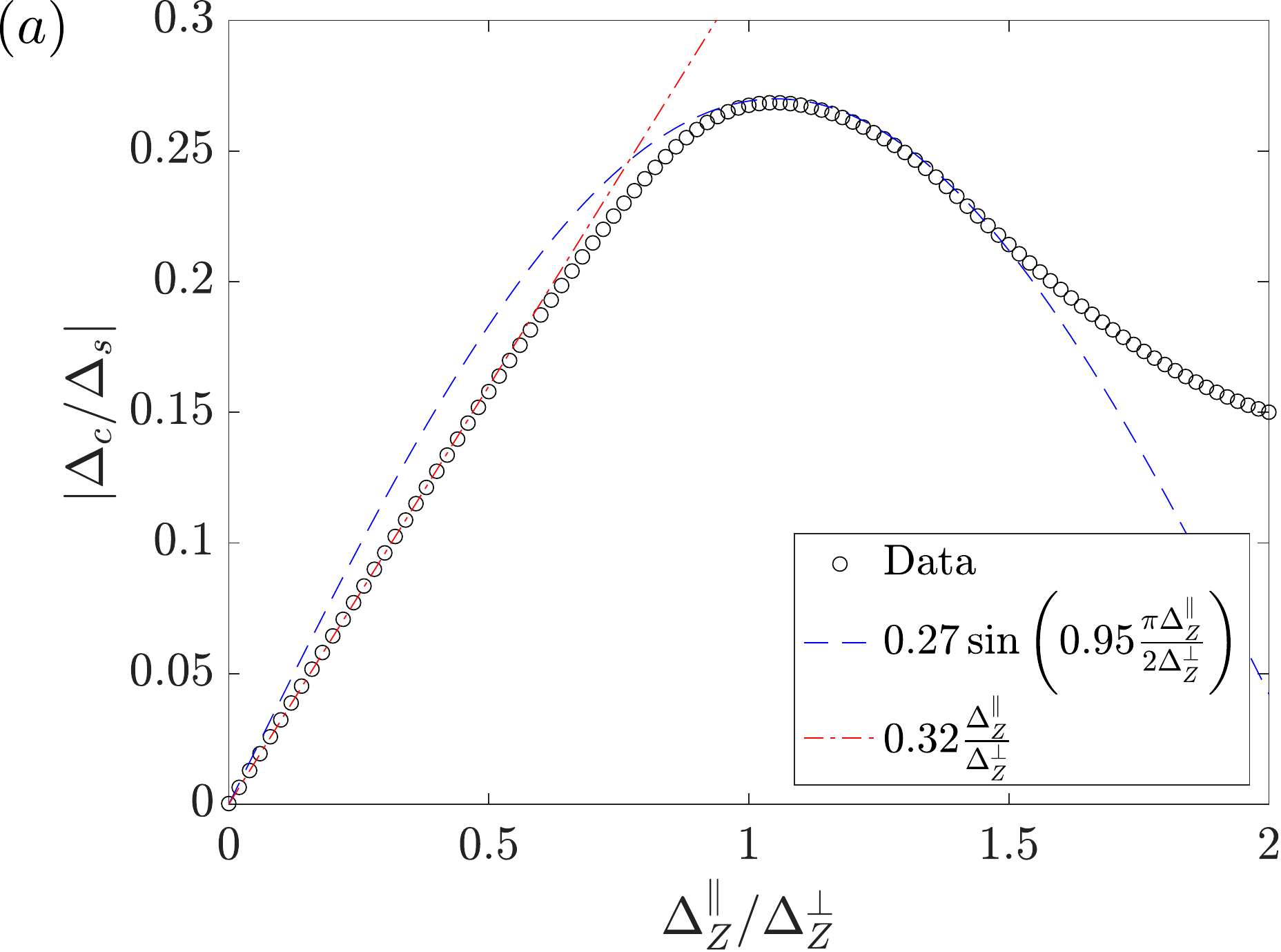}
	\caption{\label{GapInPlaneBSinus} The absolute value of the proximity-induced CAR gap $\Delta_c$ relative to SC parent gap $\Delta_s$ as a function of the ratio of in-plane and out-of-plane components of Zeeman-energy, $\Delta_Z^\parallel/\Delta_Z^\perp$. We used the following parameters: $N_n=50$, $N_s=36$, $t_s=t_c=t$, $\mu_s=0.1t$, $\Delta_s=0.02t$, $\phi=0.01$, $\mu_n=0.0625t$, $\alpha_s=0$, and $\zeta=10a$. In this figure, we fix $\Delta_Z^\perp=0.04t$, whereas in Fig.~\ref{GapInPlaneBSinusMain} we used with a Zeeman term of constant magnitude $\Delta_Z=0.04t$ that is rotated homogeneously within the QHE regions (oppositely for the two QHE parts) from out-of-plane to in-plane directions with angle $\theta$.}
\end{figure}

Here, we provide a heuristic derivation for the dependence of the induced gap on the angle $\theta$. We denote the spin basis with quantization along the $z$ axis as $\ket{\uparrow}$ and $\ket{\downarrow}$. In the absence of an in-plane magnetic field, the spin is quantized along the same axis in the normal and in the superconducting region. For instance, we can express the spin dependence of a Cooper pair as
\begin{equation}
\ket{S}=\frac{\ket{\uparrow}\otimes\ket{\downarrow}-\ket{\downarrow}\otimes\ket{\uparrow}}{\sqrt{2}}.
\end{equation}
In presence of the in-plane component of magnetic field, the spin is not quantized anymore along this axis in the edge states and the new basis can be expressed as a rotation of the previous basis by the angle $\theta$
\begin{equation}
\ket{E_\pm}=\cos\left(\frac{\theta}{2}\right)\ket{\uparrow}\pm\sin \left(\frac{\theta}{2}\right)\ket{\downarrow},
\end{equation}
where $\pm$ indicates the two states with opposite spin polarization, i.e where the angle $\theta$ has an opposite sign.
The overlap between the spin-dependent part of edge state and Cooper pair wave-functions is 
\begin{equation}
\braket{S|E_+ E_-}=\frac{\sin\left(\theta\right)}{\sqrt{2}}.
\end{equation}
Notice that for the edge states we assumed that the spin tilting occurs with opposite angles $\theta$ and $-\theta$. This equation justifies the presence of the factor $\sin\left(\theta\right)$ in the  induced crossed Andreev gap as the overlap of the tilted edge state spin structure with the singlet CP spin wave function. 

In Fig.~\ref{GapInPlaneBSinus}, we computed numerically the dependence of $\Delta_c$ on the ratio of in-plane and out-of-plane components of Zeeman-energy, $\Delta_Z^\parallel/\Delta_Z^\perp$. In these plots, we checked that the gap starts linearly with $\Delta_Z^\parallel/\Delta_Z^\perp$ and $\theta$. In Fig.~\ref{GapInPlaneBSinus}, we also plot a sine function of $\Delta_Z^\parallel/\Delta_Z^\perp$ (blue dotted line), which fits the numerical curve up to a certain ratio and eventually deviates from it. This discrepancy shows that the exact dependence of the argument in the sine function is not linear in $\Delta_Z^\parallel/\Delta_Z^\perp$, but involves the inverse tangent function (see the definition of $\theta$). Also, the magnitude of the Zeeman field is not constant as $\Delta_Z^\parallel/\Delta_Z^\perp$ is varied, which then modifies band alignments between QHE and SC as well, so it is not anymore only the spin structure that plays a role in the magnitude of the induced crossed Andreev gap.


\subsection{Finger-width dependence for all considered cases}
\begin{figure}[]
	\includegraphics[width=\linewidth]{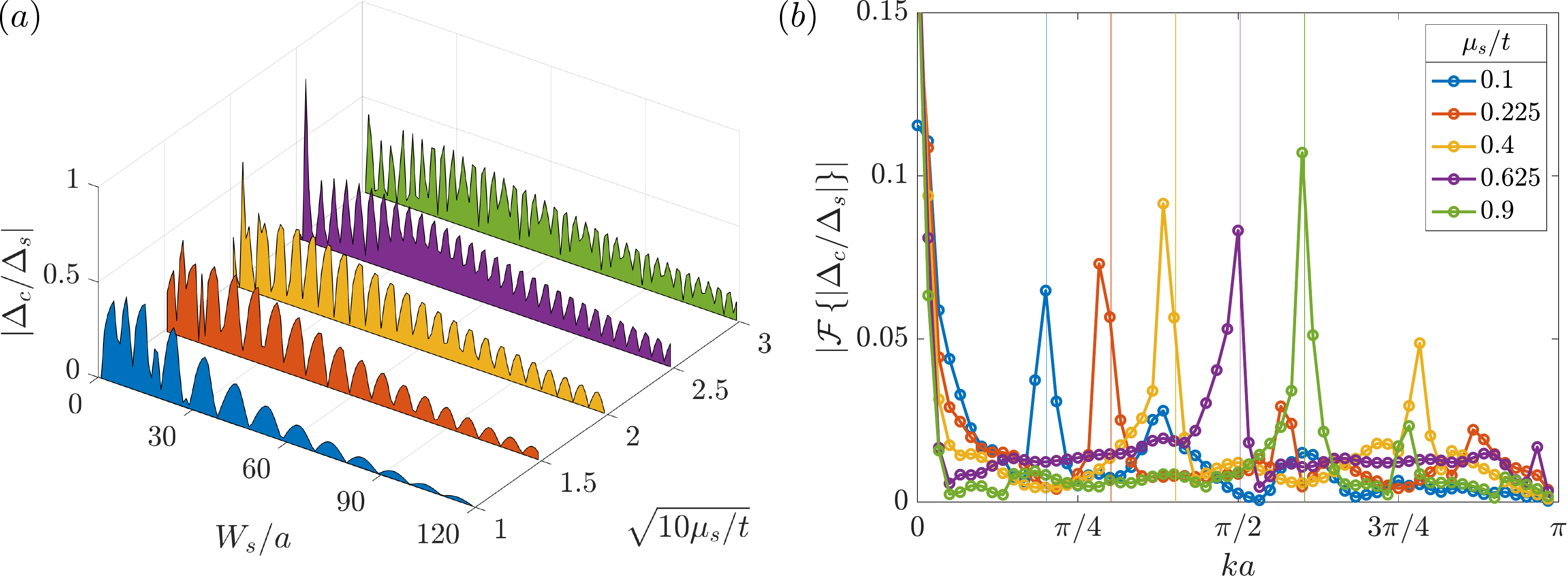}
	\caption{\label{GapOscInPlaneB} (a) Oscillation of the absolute value of the crossed gap, $\Delta_c$, induced in the $\nu=1$ QHE edge states, normalized by the superconducting gap in the SC region, $\Delta_s$, as a function of the SC region width, $W_s$, measured in units of the discretized lattice spacing $a$, for fixed in-plane magnetic field $\Delta_Z^\parallel/\Delta_Z^\perp=1$ and different values of chemical potential in the SC, $\mu_s$. (b) Fourier transformation of the data displayed in (a). The theoretically estimated Fermi momentum values are indicated by the solid colored lines. As expected, the absolute value of the gap oscillates with the period set by $2k_{F,s}$.
	Parameters of the numerics not indicated in the figure: $N_n=50$, $t_s=t_c=t$,  $\mu_n=0.0625t$, $\Delta_Z^\perp=0.04t$, $\zeta=10a$, $\Delta_s=0.02t$, $\phi=0.01$, and $\alpha_s=0$. For a better understanding, we also indicate the chemical potential dependent superconducting coherence length $\xi_s$ for all curves, $\xi_s/a\approx \left\{31,47,63,79,95\right\}$, which characterize the scale of decay for the CAR gap. We observe irregular resonances for $W_s\lesssim\xi_s$, while a smooth decaying cosine-like behavior for $W_s>\xi_s$, in line with Eqs.~\eqref{eqSM:maggap} and~\eqref{eqSM:gapOsc}. }
\end{figure}

\begin{figure}[!htbp]
	\includegraphics[width=\linewidth]{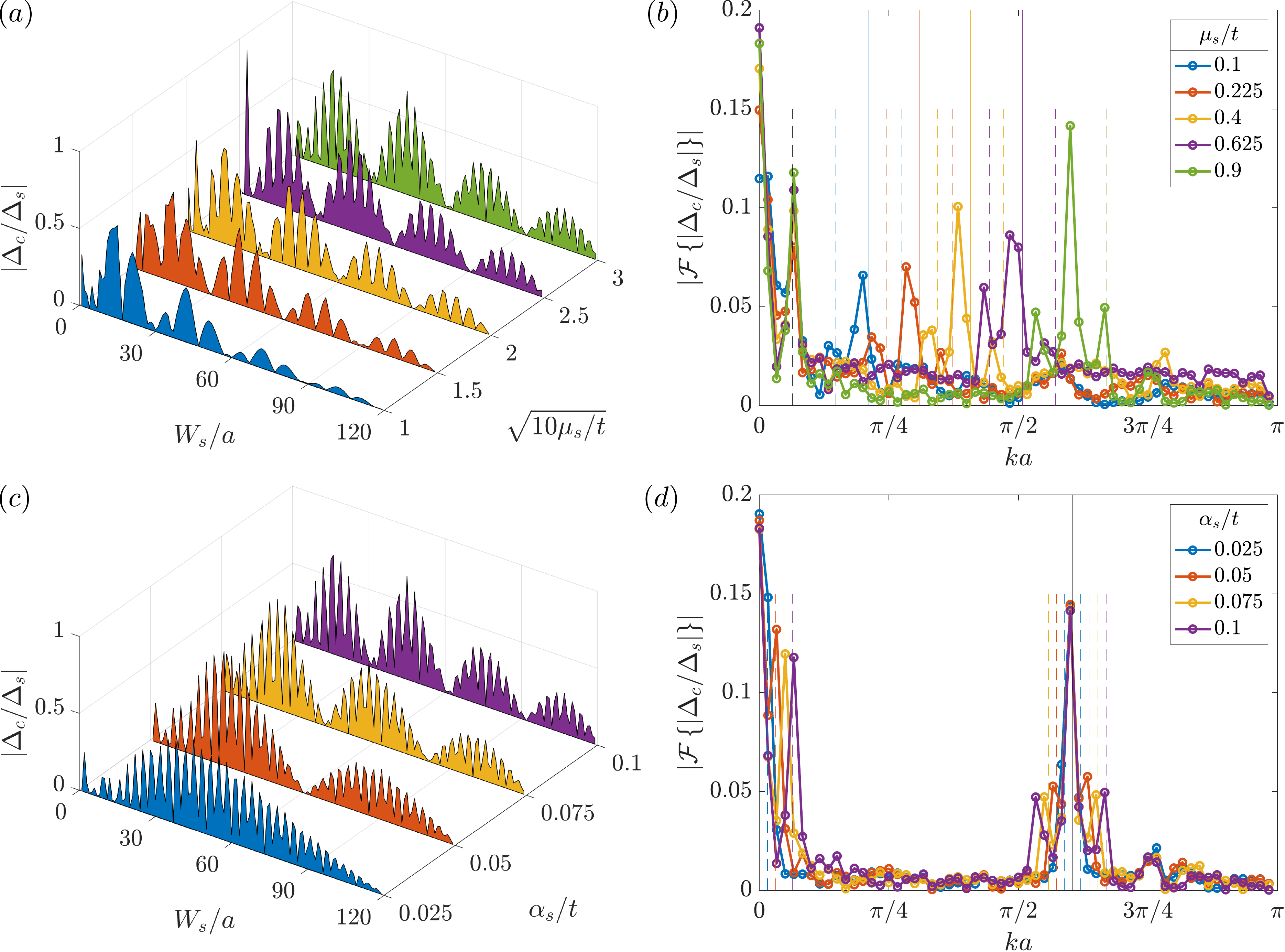}
	\caption{\label{GapOscNu1As}(a) Double oscillation of the absolute value of the crossed gap, $\Delta_c$, induced in the $\nu=1$ QHE edge states, normalized by the superconducting gap in the SC region, $\Delta_s$, as a function of the SC region width, $W_s$, measured in units of the discretized lattice spacing $a$, for fixed SOI strength $\alpha_s=0.1t$ and different values of chemical potential in the SC, $\mu_s$. (b) Fourier transformation of the data displayed in (a).  The theoretically expected Fermi-momentum values are indicated by the solid colored lines. (c) Double oscillation of the absolute value of the crossed gap, $\Delta_c$, induced in the $\nu=1$ QHE edge states, normalized by the superconducting gap in the SC region, $\Delta_s$, as a function of the SC region width, $W_s$, measured in units of the discretized lattice spacing $a$, for different values of SOI strength in the SC, $\alpha_s$, at constant SC chemical potential $\mu_s/t=0.9$. (d) Fourier transformation of the data displayed in (c). The theoretically expected fast oscillation period value is indicated by the solid black line, while dashed lines show the theoretically expected side lobe periods due to the slow envelope oscillations.  Parameters of the numerics not indicated in the figure: $N_n=50$, $t_s=t_c=t$,  $\mu_n=0.0625t$, $\Delta_Z^\perp=0.04t$, $\Delta_Z^\parallel=0$, $\zeta=10a$, $\Delta_s=0.02t$, and $\phi=0.01$.
		The superconducting coherence length $\xi_s$ for all curves in (a) takes values $\xi_s/a\approx \left\{31,47,63,79,95\right\}$, whereas in (c), it is fixed as $\xi_s/a\approx 95$. We observe irregular resonances for $W_s\lesssim\xi_s$, while a smooth, decaying double oscillatory  behavior for $W_s>\xi_s$, in line with Eqs.~\eqref{eqSM:SOIgap} and~\eqref{eqSM:gapOsc}.
	}
\end{figure}

In this section, we evaluate numerically the dependence of the induced gap on the width $W_s$ and the superconducting chemical potential $\mu_s$. We plot the $\nu=1$ case with in-plane magnetic field in Fig.~\ref{GapOscInPlaneB},  the $\nu=1$ case with SOI in Fig.~\ref{GapOscNu1As} and the $\nu=2$ case in Fig.~\ref{GapOscNu2}.

Fast oscillations in the width $W_s$ result from the existence of a definite Fermi-surface, see Figs.~\ref{GapOscInPlaneB}(a), \ref{GapOscNu1As}(a), and \ref{GapOscNu2}(a). They are characterized by a spatial period related to the magnitude of the Fermi momentum in the SC, $k_{F,s}$. In order to clear out this point, we computed the Fourier transform of the gap in each different case, see the main peaks in Figs.~\ref{GapOscInPlaneB}(b), \ref{GapOscNu1As}(b), and \ref{GapOscNu2}(b). The theoretical values correspond to the expected Fermi momentum for each different chemical potential $\mu_s$.

In Figs.~\ref{GapOscInPlaneB}, \ref{GapOscNu1As}(a), \ref{GapOscNu1As}(b), and~\ref{GapOscNu2}, one sees the fast oscillations changing with chemical potential and thus the Fermi momentum, while the envelope remains unchanged, except that the exponential suppression with the coherence length decreases as $\mu_s$ increases due to the increase of the Fermi velocity $v_{F,s}$ (and thus the coherence length $\xi_s$). One can also clearly discern in the numerically obtained plots of all three cases the resonant behavior at small $W_s\lesssim\xi_s$ described by the fraction in Eq.~\eqref{eqSM:gapOsc}.

In Figs.~\ref{GapOscNu1As}(c) and~\ref{GapOscNu1As}(d), we show an oscillating pattern of the gap as a function of SOI. Here, we can identify some slow enveloping oscillations that are induced by the spin-rotating effect of SOI. They are peculiar to the presence of the SOI that enables the overlap of the spin-polarized $\nu=1$ edge states with the singlet structure of the SC. The rotation and the overlap size are dependent on the amount of rotation that is achieved across the SC finger width of size $W_s$, thus inducing an oscillating pattern. Side peaks in the FT correspond to these oscillations dependent on the SOI parameter $\alpha_s$.


\subsection{Dependence on coupling $t_c$}
\begin{figure}[]
	\includegraphics[width=\linewidth]{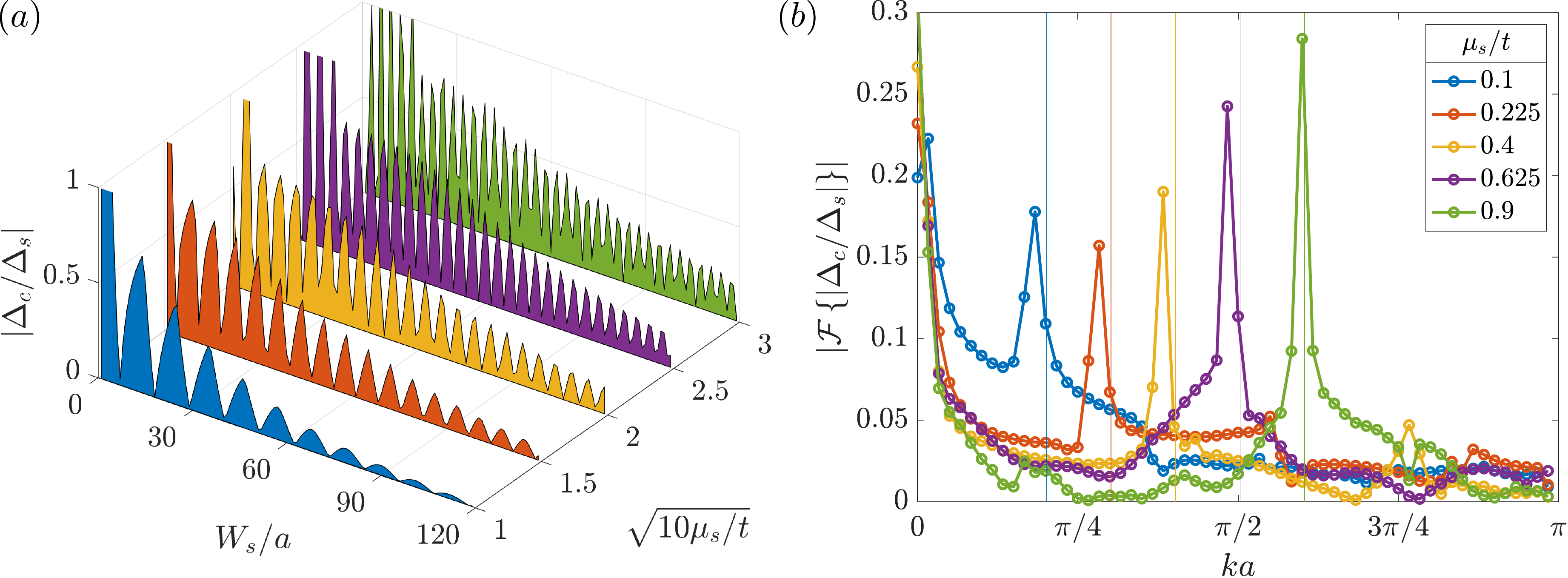}
	\caption{\label{GapOscNu2} (a) Oscillation of the absolute value of the crossed gap, $\Delta_c$, induced in the $\nu=2$ QHE edge states, normalized by the superconducting gap in the SC region, $\Delta_s$, as a function of the SC region width, $W_s$, measured in units of the discretized lattice spacing $a$, for different values of chemical potential in the SC, $\mu_s$. (b) Fourier transformation of the data displayed in the panel (a). The theoretically expected Fermi-momentum values are indicated by the solid colored lines. Parameters of the numerics not indicated in the figure: $N_n=50$, $t_s=t_c=t$,  $\mu_n=0.1025t$, $\Delta_Z^\parallel=\Delta_Z^\perp=0$, $\zeta=10a$, $\alpha_s=0$, $\Delta_s=0.02t$, and $\phi=0.01$. Superconducting coherence length $\xi_s$ for all curves reads $\xi_s/a\approx \left\{31,47,63,79,95\right\}$. We observe irregular resonances for $W_s\lesssim\xi_s$, while a smooth, decaying cosine-like behavior for $W_s>\xi_s$, in line with Eqs.~\eqref{eqSM:maggap} and~\eqref{eqSM:gapOsc}.
	}
\end{figure}
\begin{figure}[bht!]
	\includegraphics[width=\linewidth]{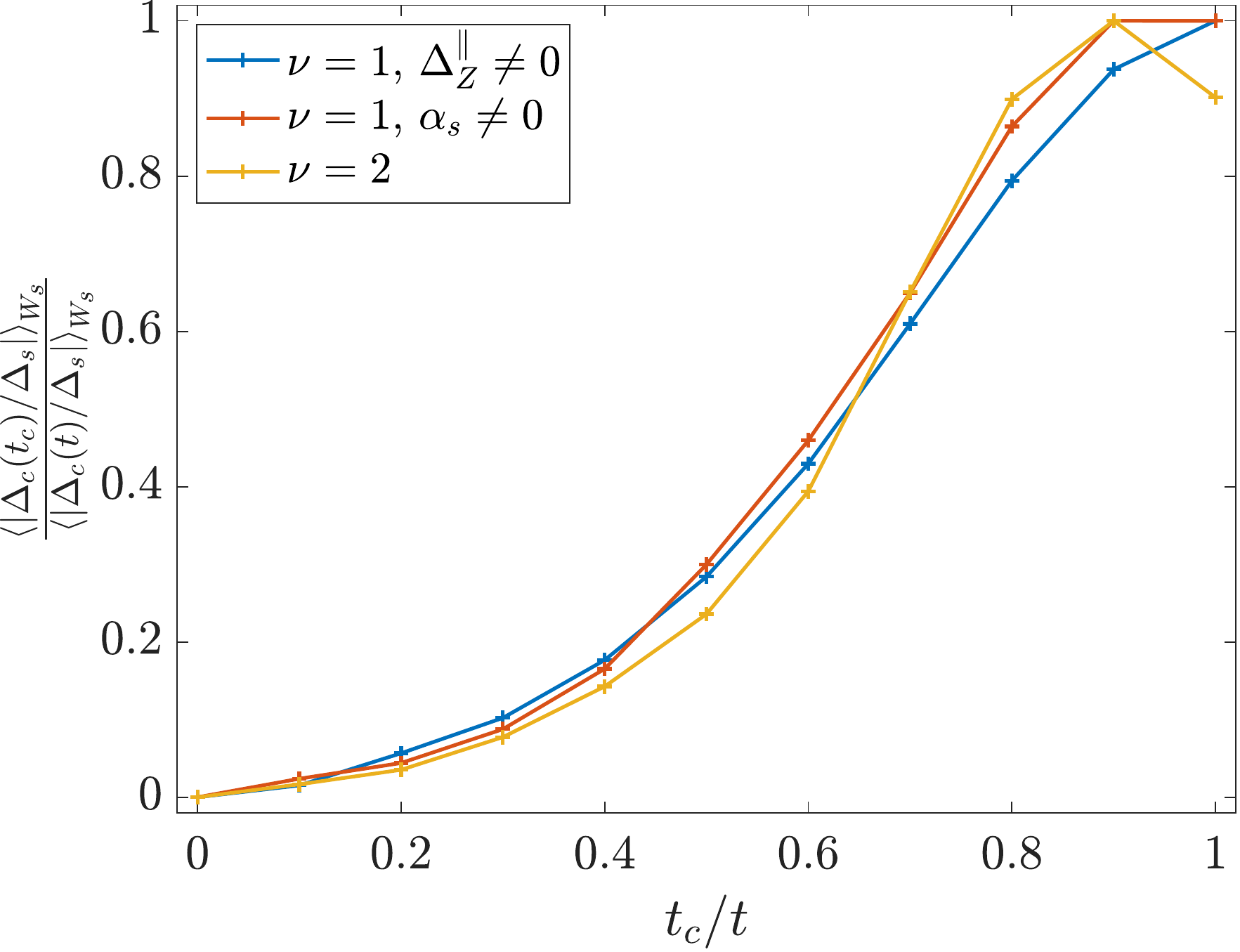}
	\caption{\label{AvgTc} Plot of the absolute value of the proximity-induced CAR gap $\Delta_c$ relative to SC parent gap $\Delta_s$, averaged over different values of $W_s$ as a function of $t_c/t$, normalized with respect to the same quantity at $t_c=t$. Curves are obtained numerically in an infinite QHE-SC-QHE system with $N_n=50$, $N_s=2,\ldots, 120$, $t_s=t$, $\mu_s=0.4t$, $\Delta_s=0.02t$, $\phi=0.01$, $\zeta=10a$, and for $\nu=1$, $\Delta_Z^\parallel\neq 0$ with $\mu_n=0.0625t$, $\Delta_Z^\parallel=\Delta_Z^\perp=0.04t$, $\alpha_s=0$, for $\nu=1$, $\alpha_s\neq 0$ with $\mu_n=0.0625t$, $\Delta_Z^\parallel=0$, $\Delta_Z^\perp=0.04t$, $\alpha_s=0.1t$, and finally for $\nu=2$ with $\mu_n=0.1025t$, $\Delta_Z^\parallel=\Delta_Z^\perp=0$, and $\alpha_s=0$ (no SOI).}
\end{figure}

Here, we evaluate numerically the dependence of the induced gap on the normal-superconductor coupling $t_c$. In Fig.~\ref{AvgTc}, we display the ratio $\Delta_c/\Delta_s$ averaged over different values of $W_s$ in the same range as shown in Figs.~\ref{GapOscInPlaneB}--\ref{GapOscNu2}, as a function of $t_c/t$, normalized with respect to the same quantity at $t_c=t$, for all three different cases considered. First of all, we do find only a weak dependence on the filling factor, on SOI, or on in-plane magnetic field to a large degree.  One can see that the induced gap start to develop quadratically and reach $1$ only close to $t_c=t$. These results are in accordance with Eqs.~\eqref{eqSM:maggap}--\eqref{eqSM:nu2}. 
 
\subsection{Stability against disorder}
In this part, we test the stability of the induced gap with respect to disorder in the SC chemical potential and in the interface coupling $t_c$. The latter type of disorder can arise due to an imperfect QHE-SC coupling or due to the presence of defects at the interface. 

\begin{figure}[b!]
	\includegraphics[width=\linewidth]{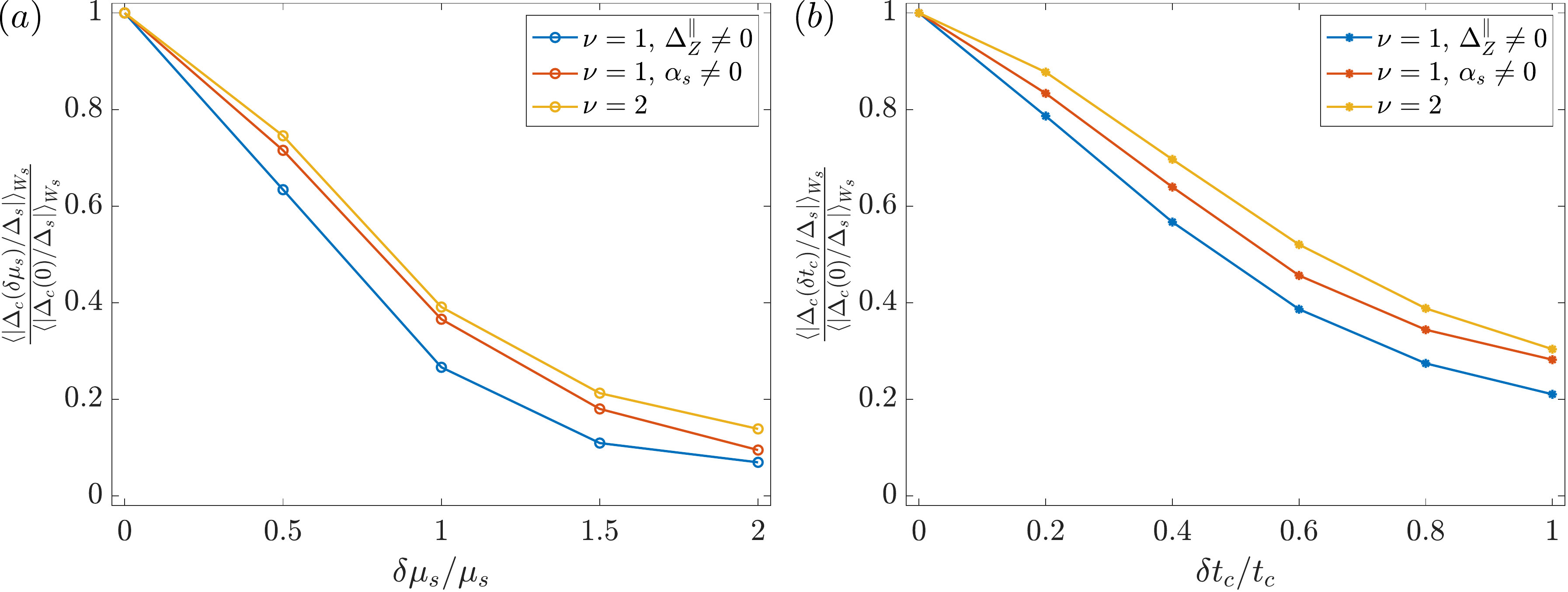}
	\caption{\label{AvgDmusDtc} (a) The absolute value of the proximity-induced CAR gap $\Delta_c$ relative to SC parent gap $\Delta_s$ averaged over different values of $W_s$ with respect to the disorder  (a) in the SC chemical potential $\mu_s$ and (b) in the interface coupling $t_c$, normalized with respect to the same quantity without disorder. Curves are obtained with the same respective parameters as in Fig.~\ref{AvgTc}, by numerical exact diagonalization of a long but finite QHE-SC-QHE planar junction.}
\end{figure}
In Fig.~\ref{AvgDmusDtc}, we computed the ratio $\Delta_c/\Delta_s$ averaged over different values of $W_s$ in the same range as in Fig.~\ref{GapOscInPlaneB} for these two types of disorder, normalized with respect to the same quantity without disorder. First of all, we note that there is no striking difference between the three different cases considered. The important point to notice is that, even though the presence of disorder causes a reduction of the induced gap, the gap keeps a finite value even for disorder size of the order of $\mu_s$ [in Fig.~\ref{AvgDmusDtc} (a)] or $t_c$ [in Fig.~\ref{AvgDmusDtc} (b)]. 
For Fig.~\ref{AvgDmusDtc}(a), the SC chemical potential is  redefined  as  $\mu_{s, lm}=\mu_s+\delta\mu_s u_{lm}$, where $u_{lm}\sim U(-1,1)$ are iid uniform random variables on each SC site. Similarly in Fig.~\ref{AvgDmusDtc}(b), the interface disorder varies the coupling  strength as $t_{s, lm}=t_c+\delta t_c\, u_{lm}$, sampled independently for each interface link. 


\section{Transport calculation details \label{app:TransportCalc}}
If one symmetrizes the Hamiltonian of Eq.~(\ref{eq:fingerH}) in particle-hole space, that is, writes it in the full BdG form, it can be represented as a rather sparse matrix $\mathbb{H}$ of dimension $4N_xN_y\times 4N_xN_y$ corresponding to site-, spin-, and electron-hole degrees of freedom. Its elements are addressed as $\left[\mathbb{H}\right]_{lm\sigma\tau,l'm'\sigma'\tau'}$.
One defines then the retarded and advanced Green's functions (GFs) of the system at energy $E$ as
\begin{equation}
\mathbb{G}^{R/A}(E)=\left[(E\pm i \eta)\mathbb{I}-\mathbb{H}-\Sigma^{R/A}(E)\right]^{-1},
\end{equation}
where $\eta$ is the infinitesimal (small) linewidth, $\mathbb{I}$ is the identity matrix of the appropriate dimension and $\Sigma^{R/A}(E)$ describe the self-energy at the same frequency of anything attached to the principal region, in particular,
\begin{align}
&\left[\Sigma^{R/A}\right]_{1m\sigma\tau,1m'\sigma'\tau'}=\left[\Sigma_L^{R/A}\right]_{m\sigma\tau,m'\sigma'\tau'},
\\
&\left[\Sigma^{R/A}\right]_{N_x m\sigma\tau,N_xm'\sigma'\tau'}=\left[\Sigma_R^{R/A}\right]_{m\sigma\tau,m'\sigma'\tau'}
\end{align}
describe the self-energies of the left- and right leads, $4N_y\times 4N_y$ dimensional matrices themselves, which can be easily obtained trough an iterative calculation of surface GFs~\cite{Lee1981,Sancho1984,Sancho1985}. 
Elements of the full GF can be calculated efficiently by the recursive GF method~\cite{MacKinnon1985}.
We note that $\mathbb{G}^A(E)=\left[\mathbb{G}^R(E)\right]^\dagger$ and therefore define $\mathbb{G}=\mathbb{G}^R(E)$ for the sake of simplified notation. In addition, we define the broadening functions
\begin{equation}
\Gamma_{L/R}^\tau=i\left(\Sigma_{L/R}^R-\Sigma_{L/R}^A\right)_{\tau\tau},
\end{equation}
where we already separate out the electron-hole space dependence, which is --- since the leads are normal semi-infinite QHE systems --- trivially diagonal anyway. $\Gamma_{L/R}^\tau$ are now $2N_y\times2N_y$ square matrices.

Firstly, to calculate the (spatially dependent) LDOS at energy $E$, one needs the diagonal elements of the GF:
\begin{equation}
LDOS(E,l,m)=-\frac{1}{2\pi}\mathrm{Im}\left[\mathrm{Tr}\ \mathbb{G}_{lm,lm}(E)\right],
\end{equation}
where the trace is taken over the spin- and particle-hole space and the unusual factor of  2 accounts for the base doubling in the BdG formalism.

Secondly, to study transport, we need the following GF elements:
\begin{align}
\left[G_{LL}^{\tau\tau'}\right]_{m\sigma,m'\sigma'}&=\left[\mathbb{G}\right]_{1m\sigma\tau,1m'\sigma'\tau'},\\
\left[G_{RL}^{\tau\tau'}\right]_{m\sigma,m'\sigma'}&=\left[\mathbb{G}\right]_{N_x m\sigma\tau,1m'\sigma'\tau'},\\
\left[G_{LR}^{\tau\tau'}\right]_{m\sigma,m'\sigma'}&=\left[\mathbb{G}\right]_{1m\sigma\tau,N_x m'\sigma'\tau'},\\
\left[G_{RR}^{\tau\tau'}\right]_{m\sigma,m'\sigma'}&=\left[\mathbb{G}\right]_{N_x m\sigma\tau,N_x m'\sigma'\tau'},
\end{align}
where again we separate out in the definition the electron-hole space indices to facilitate later discussion of normal- (diagonal in the e-h space) and anomalous or Andreev processes (off-diagonal in e-h). These $G_{ij}^{\tau\tau'}$ GFs are again matrices of $2N_y\times2N_y$ dimension, indices take values as $i,j=\{L,R\}$ and $\tau,\tau'=\{e,h\}$.
Then one defines the energy dependent transport coefficients~\cite{Meir1992} as
\begin{equation}
T_{ij}^{\tau\tau'}(E)=\mathrm{Tr}\left\{\Gamma_i^\tau(E)G_{ij}^{\tau\tau'}(E)\Gamma_j^{\tau'}(E)\left[G_{ij}^{\tau\tau'}(E)\right]^\dagger\right\},
\end{equation}
where trace is over the $y$ direction sites and over spin, yielding a single number, which describes the energy-dependent normal ($\tau=\tau'$) and Andreev ($\tau\neq\tau'$)  reflection ($i=j$) and transmission ($i\neq j$) probabilities in the principal region of  the system, as Fig.~\ref{fingersetup} shows.

The transport setup in Fig.~3. of the main text represents a multi-terminal junction with three terminals or leads: terminal $L$, where the generated current $I$ is fed in; the grounded SC finger, terminal $S$, which also drains the current; as well as terminal $R$, which is a floating gate only to measure the downstream resistance. The multi-terminal current-bias relations in the linear response Landauer-B\"uttiker formalism read~\cite{Lambert1998,Beconcini2018}
\begin{equation}
I_i=\sum_j a_{ij}\left(V_j-V\right),
\end{equation}
where $I_i$ is the current from lead $i$, $V_j$ is the potential of lead $j$, $V=V_S=\mu_s/e$ is the electrochemical potential of the SC condensate around which one linearizes for small bias differences between leads (meaning that the up- and downstream resistance measurement needs to happen with weak driving current $I$), and finally the proportionality coefficients are
\begin{multline}
a_{ij}=\frac{e^2}{h}\int_{-\infty}^{\infty}\ dE \left[-\frac{\partial f(E)}{\partial E}\right]\\ \times\left[N_i^e(E)\delta_{ij}-T_{ij}^{ee}(E)+T_{ij}^{eh}(E)\right].
\end{multline}
Here a factor of 2 is omitted compared to Ref.~\onlinecite{Lambert1998} because our transmission coefficients include contributions of both spin species. Also, the above form explicitly takes into account the particle-hole symmetry of our system. $f(E)=\left[1+\exp\left(E/k_B T\right)\right]^{-1}$ is the Fermi function at temperature $T$, and $N_i^e(E)=\sum_{j,\tau}T_{ij}^{e\tau}(E)$ counts the number of open channels for electron-like excitations in lead $i$.

To express the resistances $R_{U/D}=V_{U/D}/I$ for  $V_U=V_L-V_S$, $V_D=V_R-V_S$ and conditions $I_L=I$ and  $I_R=0$, we can write 
\begin{gather}
a_{LL}V_U+a_{LR}V_D =I,\\ a_{RL}V_U+a_{RR}V_D=0,\\
R_U=\frac{a_{RR}}{a_{LL}a_{RR}-a_{LR}a_{RL}}, \\ R_D=-\frac{a_{RL}}{a_{LL}a_{RR}-a_{LR}a_{RL}},\\
a_{LL}=\frac{e^2}{h}\left(T_{LR}^{ee}+T_{LR}^{eh}+2T_{LL}^{eh}\right),\\
a_{LR}=\frac{e^2}{h}\left(T_{LR}^{eh}-T_{LR}^{ee}\right),\\a_{RR}=\frac{e^2}{h}\left(T_{RL}^{ee}+T_{RL}^{eh}+2T_{RR}^{eh}\right)
,\\
a_{RL}=\frac{e^2}{h}\left(T_{RL}^{eh}-T_{RL}^{ee}\right),
\end{gather}
where the transmission coefficients are evaluated (for a given temperature) as
\begin{equation}
T_{ij}^{e\tau}=\int_{-\infty}^{\infty}\ dE\frac{T_{ij}^{ e\tau}(E)}{4k_B T \cosh^2\left(E/2k_B T\right)},
\end{equation}
finally yielding
\begin{gather}
R_U=R_Q\frac{T_{RL}^{ee}+T_{RL}^{eh}+2T_{RR}^{eh}}{D},\\ R_D=R_Q\frac{T_{RL}^{ee}-T_{RL}^{eh}}{D},\\
\begin{aligned}
D&=T_{LR}^{ee}T_{RL}^{eh}+T_{RL}^{ee}T_{LR}^{eh}+T_{LL}^{eh}\big(T_{RL}^{ee}+T_{RL}^{eh}\\&+T_{RR}^{eh}\big)+T_{RR}^{eh}\left(T_{LR}^{ee}+T_{LR}^{eh}+T_{LL}^{eh}\right)\end{aligned}
\end{gather}
in line with results of Ref.~\onlinecite{Beconcini2018} if adapted to the $\nu=1$ filling case. Here, $R_Q=h/2e^2$, as defined in the main text, along with $T_{RL}^{ee}=T^N$ and $T_{RL}^{eh}=T^A$, the normal- and Andreev transmission probabilities. Due to the chiral nature of edge states, there is no backscattering (normal or Andreev) in the system, meaning $T_{ii}^{\tau\tau'}=0$. A further consequence of the chirality and of the fact that the SC finger is only in contact with the lower edge mode is $T_{LR}^{eh}=0$, because only the SC can convert electrons into holes. This entails, lacking backscattering, that $T_{LR}^{ee}=\nu=1$, i.e., the upper edge perfectly transmits electrons from right to left. In the lower edge an electron is either transmitted from left to right as an electron or a hole (leaving a CP in the SC), meaning $T^N+T^A=\nu=1$, as stated in the main text. All these simplifications yield $D=\nu T^A$ and with that $R_{U/D}=R_Q\left(T^N\pm T^A\right)/\left(\nu T^A\right)$. Note that measuring the difference $R_U-R_D=h/\left(\nu e^2\right)$ is equivalent to measuring the nonlocal or Hall resistance in the QHE system.

A remark is due in order to explain the $T^A\to 0$ limit of the above expressions of the resistances, which diverge in this case. This limit means that the QHE edges transmit electrons without talking to the SC finger. As the SC is the only drain on the circuit (lead $R$ being a floating voltage probe), all pumped-in current from lead $L$ has to leave the QHE edge circle again through the same lead, so effectively it is impossible to push any net current into the system, which is exactly the case of infinite resistance.

\begin{figure}[t]
	\includegraphics[width=\linewidth]{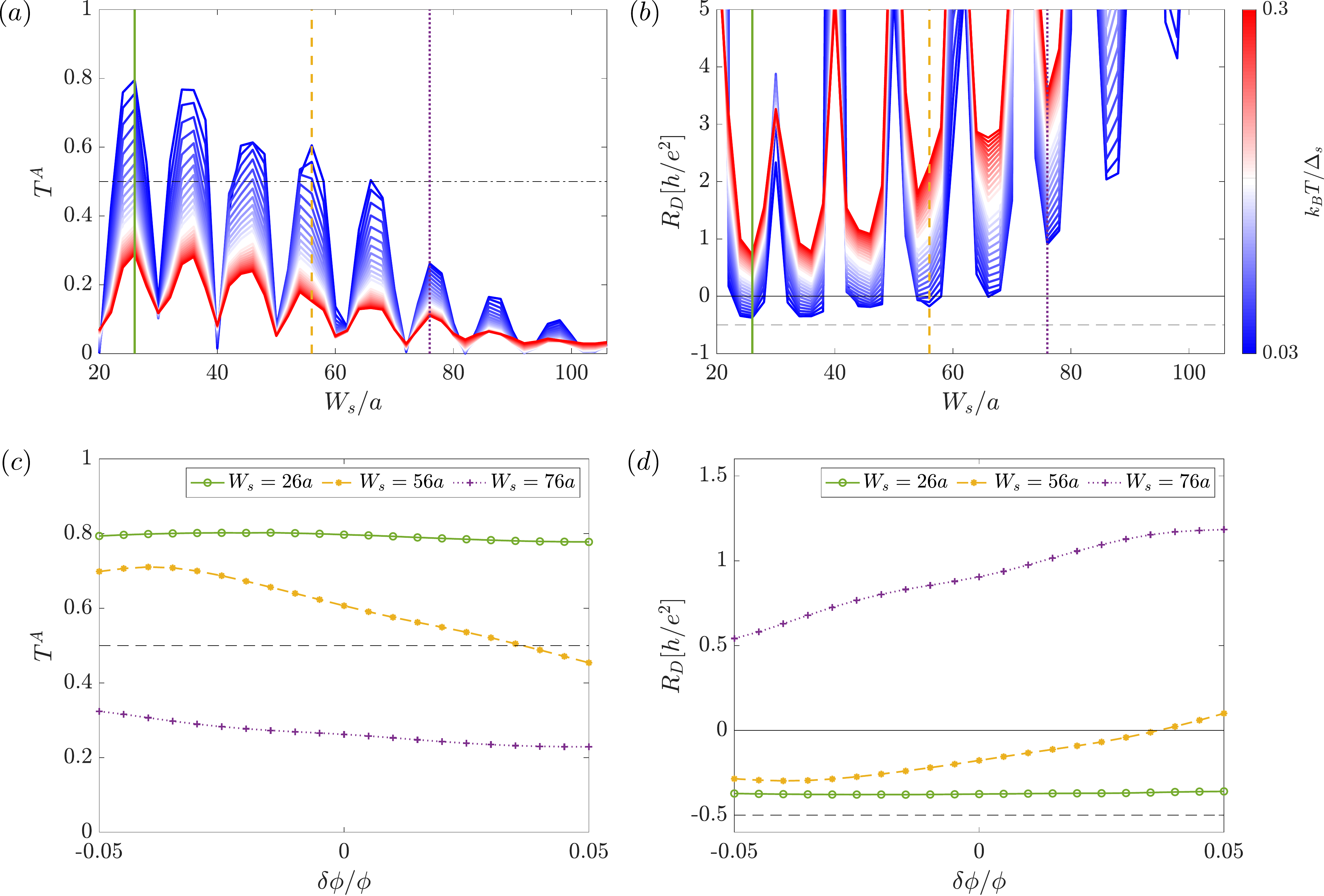}
	\caption{\label{RdNu1InPlaneB} (a) Andreev transmission coefficient between the two leads, $T^A$, as a function of the finger width $W_s$ at different temperatures $T$ for the in-plane magnetic field case.
		The dash-dotted line at $0.5$ indicates the threshold above which values of anomalous transmission negative downstream resistances are observed. (b) Calculated width dependence of downstream ($R_D$) resistances. The dashed line indicates the theoretically achievable minimum resistance, $R_{D}=- R_Q$ for $\nu=1$ filling. (c) Stability of the Andreev transmission coefficient $T^A$ as a function of the magnetic flux variation $\delta\phi$.
		(d) Stability of the downstream resistance $R_D$ as a function of the magnetic flux variation $\delta\phi$. This magnetic flux controls the orbital effect in the QHE region. The stability is checked for different widths associated with peaks in the transmission. In correspondence with the chosen values of $W_s$, lines are drawn in (a) and (b) with the same color code as in (c) and (d).
		Parameters: $N_x=150$, $N_y=120$, $S=90$, $t_s=t_c=t$, $\mu_s=0.1t$, $\Delta_s=0.02t$, $\phi=0.01$, $\mu_n=0.0625t$, $\Delta_Z^\perp=0.04t$, $\Delta_Z^\parallel=0.02t$, $\zeta=10a$, and $\alpha_s=0$.}
\end{figure}


\section{Transport setup \label{app:TransportResults}}
In this section, we consider the transport setup with a finite-size SC finger of length $L_s$. Firstly, we check the stability of negative resistance peaks when the filling factor is changed, pointing out the differences between the case at $\nu=1$, where CAR dominates, and the case at $\nu=2$, where local Andreev reflection (LAR) processes are present. Secondly, we show the zero-energy local density of states (LDOS) for disordered systems, in order to prove the stability of zero-energy peaks and their compatibility with emergence of Majorana bound states. Lastly, we present a heuristic explanation outlining the relationship between localized MBSs and the Andreev transmission coefficient.

\begin{figure}[!tbp]
	\includegraphics[width=\linewidth]{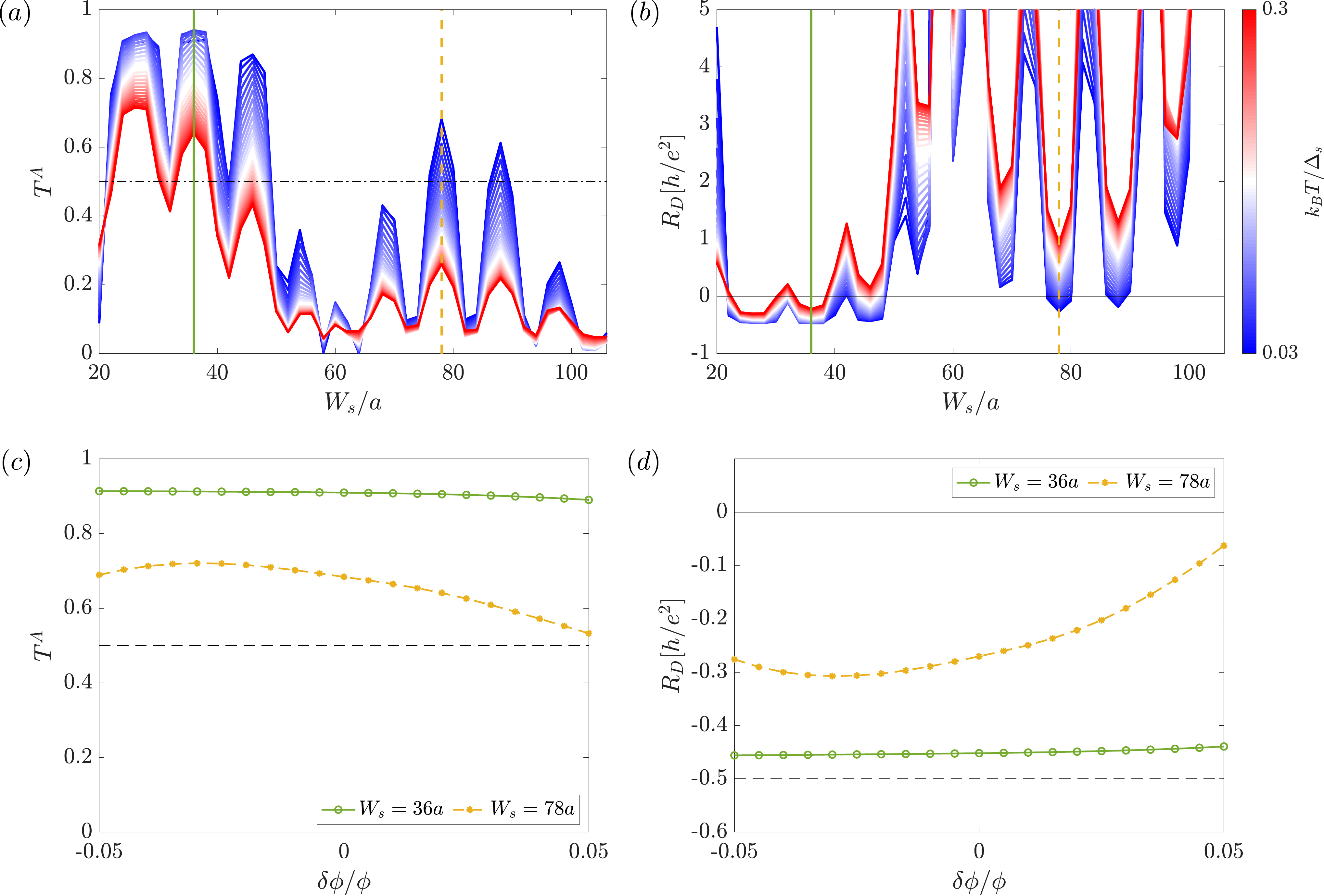}
	\caption{\label{RdNu1As} (a) Andreev transmission coefficient between the two leads, $T^A$, as a function of the finger width $W_s$ at different temperatures $T$ for the SOI case.
		The dash-dotted line at $0.5$ indicates the threshold above which values of anomalous transmission negative downstream resistances are observed. (b) Calculated width dependence of downstream ($R_D$) resistances. The dashed line indicates the theoretically achievable minimum resistance, $R_{D}=- R_Q$ for $\nu=1$ filling. (c) Stability of the Andreev transmission coefficient $T^A$ as a function of the magnetic flux variation $\delta\phi$.
		(d) Stability of the downstream resistance $R_D$ as a function of the magnetic flux variation $\delta\phi$. This magnetic flux controls the orbital effect in the QHE region. The stability is checked for different widths associated with peaks in the transmission. In correspondence with the chosen values of $W_s$, lines are drawn in (a) and (b) with the same color code as in (c) and (d).	Note how the double oscillations of $T^A$ in (a) are reminiscent of the double oscillations of the induced CAR gap displayed for the $\nu=1$ SOI case in Fig.~\ref{GapOscNu1As}. Parameters: $N_x=150$, $N_y=120$, $S=90$, $t_s=t_c=t$, $\mu_s=0.1t$, $\Delta_s=0.02t$, $\phi=0.01$, $\mu_n=0.0625t$, $\Delta_Z^\perp=0.04t$, $\Delta_Z^\parallel=0$, $\zeta=10a$, and $\alpha_s=0.05t$.}
\end{figure}


\subsection{Finger-width dependence of downstream resistance and stability of negative resistance dips against magnetic flux variation (orbital effects)}
\begin{figure}[!]
	\includegraphics[width=\linewidth]{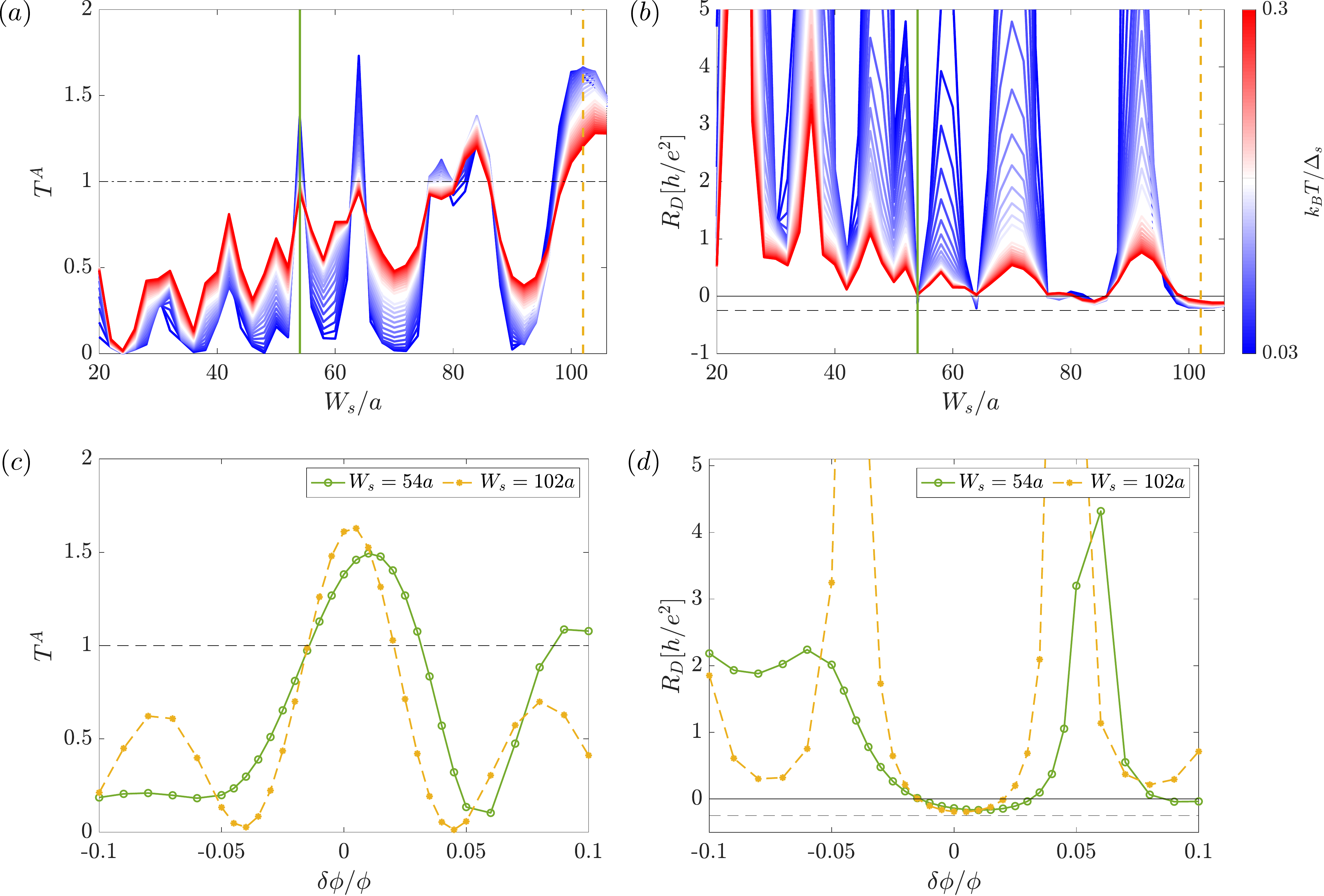}
	\caption{\label{RdNu2}(a) Andreev transmission coefficient between the two leads, $T^A$, as a function of the finger width $W_s$ at different temperatures $T$ for the $\nu=2$  case. 
		The dash-dotted line at $1$ indicates the threshold above which values of anomalous transmission negative downstream resistances are observed. (b) Calculated width dependence of downstream ($R_D$) resistances. The dashed line indicates the theoretically achievable minimum resistance, $R_{D}=- R_Q/2$ for $\nu=2$ filling. (c) Stability of the Andreev transmission coefficient $T^A$ as a function of the magnetic flux variation $\delta\phi$.
		(d) Stability of the downstream resistance $R_D$ as a function of the magnetic flux variation $\delta\phi$. This magnetic flux controls the orbital effect in the QHE region. The stability is checked for different widths associated with peaks in the transmission. In correspondence with the chosen values of $W_s$, lines are drawn in (a) and (b) with the same color code as in (c) and (d).
		Parameters: $N_x=150$, $N_y=120$, $S=90$, $t_s=t_c=t$, $\mu_s=0.1t$, $\Delta_s=0.02t$, $\phi=0.01$, $\mu_n=0.1025t$, $\Delta_Z^\perp=\Delta_Z^\parallel=0$, $\zeta=10a$, and $\alpha_s=0$.}
\end{figure}

In Figs.~\ref{RdNu1InPlaneB},~\ref{RdNu1As}, and~\ref{RdNu2}, we plot the Andreev transmission coefficient between the two leads, $T^A$,  and the downstream resistance $R_D$, checking their stability with respect to magnetic flux variation, for all the cases considered. The magnetic flux controls the orbital effect in the QHE region. In each plot, panels (a) and (b) show the Andreev transmission coefficient and the downstream resistance, respectively, as a function of the width $W_s$, while panels (c) and (d) present the stability of the same quantities with respect to variation $\delta\phi$ of the magnetic flux, $\phi=\phi+\delta\phi$.
The main goal is to point out the difference between the case at $\nu=1$, where only nonlocal Andreev reflections are possible, and the case at $\nu=2$, where also local Andreev reflection (LAR) processes are present. In the spin-unpolarized edge states at $\nu=2$, the negative resistance dips can originate from the formation of Andreev edge states (AES), which can be interpreted as an alternating skipping orbit of electrons and holes locally Andreev reflected at the SC interface. Since the radius of these skipping orbits depend on the magnetic length $\ell_B$, negative resistance dips induced by AES should display an oscillating pattern in the magnetic field. The formation of AES is ruled out by the spin-polarization of $\nu=1$ edge states at the two separate QHE-SC interfaces. We also check that peaks and dips at $\nu=1$ are not oscillating against the variation of magnetic field, while they exhibit an oscillating pattern in the variation of magnetic flux at $\nu=2$.


\subsubsection{$\nu=1$ case, in-plane magnetic field mechanism}
In Figs.~\ref{RdNu1InPlaneB}, Andreev transmission and the associated downstream resistance remains very stable with respect to flux variations. 


\subsubsection{$\nu=1$ case, spin-orbit interaction mechanism}
For the spin-orbit case (see Fig.~\ref{RdNu1As}), we observe the same behavior as for the previous case, thus showing that the negative resistance peaks at $\nu=1$ are independent of the magnetic length. We confirm that, at $\nu=1$, we can rule out the explanation for the negative resistance peaks based on AES.

\begin{figure}[!tbp]
	\includegraphics[width=\linewidth]{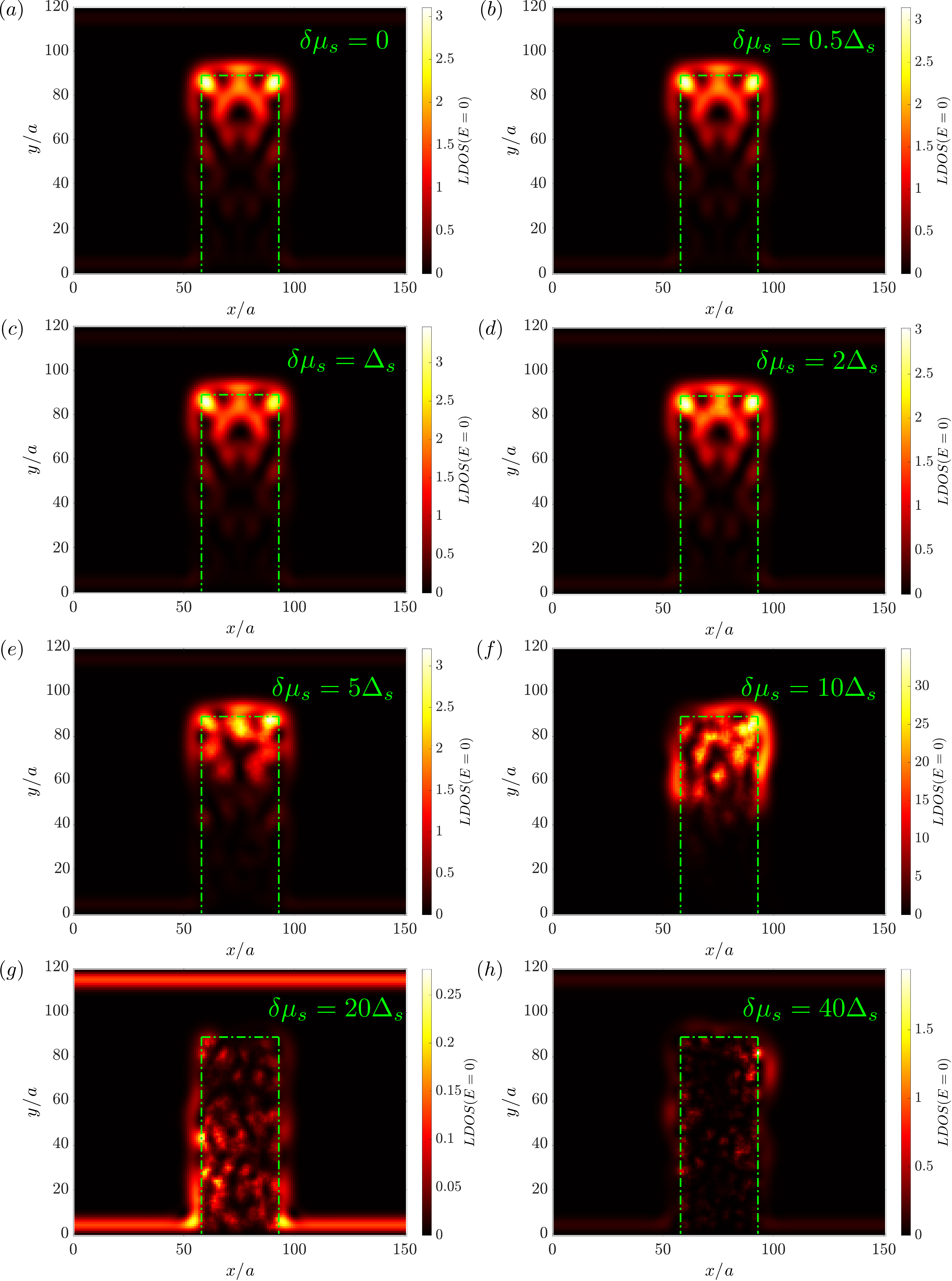}
	\caption{\label{LDOS}(a) LDOS at zero energy ($E=0$),
		in the finite-geometry SC finger setup. The green dash-dotted lines indicate the SC finger. Note that
		edge states along the SC finger gap out and a localized zero-energy MBS forms on the tip of the finger, as well as its delocalized partner in the lower edge states (away from the finger), invisible at the scale adapted to the MBS density. The strength of disorder is (a) $\delta\mu_s=0$, (b) $\delta\mu_s=0.5\Delta_s$, (c)$\delta\mu_s=\Delta_s$,  (d) $\delta\mu_s=2\Delta_s$, (e) $\delta\mu_s=5\Delta_s$, (f) $\delta\mu_s=10\Delta_s$, (g) $\delta\mu_s=20\Delta_s$, and (h) $\delta\mu_s=40\Delta_s$. Apart from the presence of disorder, densities are obtained numerically with the same parameters as in Fig. 4.(a) in the main text.}
\end{figure}
\subsubsection{$\nu=2$ case}

In Fig.~\ref{RdNu2}, the negative resistance peaks are significantly dependent on the orbital effects and exhibit an oscillating pattern with respect to the flux variation $\delta\phi$, for different values of the finger width. For spin-unpolarized edge states at $\nu=2$, we interpret the appearance of negative resistance values as a result of LAR processes and AES formation. 


\subsection{Local density of states and Majorana zero-energy modes}
In Fig.~\ref{LDOS}, we show the LDOS at zero energy for different values of disorder in the SC chemical potential. The plot in Fig.~\ref{LDOS}(a) is identical to the one in Fig. 4(a) of the main text, exhibiting a zero-energy peak at the tip of the SC finger. In panels (b)--(g), we show the same quantity for an increasing value of disorder in the SC chemical potential. Remarkably, the peaks in the zero-energy LDOS remain up to significant amounts of disorder, i.e., $\delta\mu_s\approx20 \Delta_s$. We consistently observe the stability of zero-energy peaks in correspondence with negative values of downstream resistance. This result is in agreement with the hypothesis of the emergence of Majorana bound states in our setup. The second zero-energy peak at the bottom of the SC finger hybridizes with the continuum of QHE edge states and cannot be observed in the zero-energy LDOS. This hybridization is key to the efficient electron-to-hole conversion in the transport and thus to the occurrence of negative downstream resistances.


\begin{thebibliography}{113}
\bibitem [{FN2()}]{FN2}%
	\BibitemOpen
	\href@noop {} {\bibinfo  {journal} {The form of exponential decay is merely a
			modeling choice, in reality the decay would be a power-law, but this does not
			change our results qualitatively. Important for our purposes is to have the
			decay length $\zeta$ long enough to provide significant in-plane components
			in the chiral edge modes, that is $\zeta>\ell_B$, with the magnetic length
			$\ell_B=\sqrt{\hbar/eB_0}$}\ }\BibitemShut {NoStop}%
\bibitem [{FN1()}]{FN1}%
	\BibitemOpen
	\href@noop {} {\bibinfo  {journal} {The exchange-coupled spins of the chiral
			edge mode around the finger will align with the local external field if the
			finger is long enough for the accumulated Zeeman-energy on the sides to win
			over the exchange energy excess on the tip due to misaligned spins that have
			to turn between the two sides}\ }\BibitemShut {NoStop}%
\end{thebibliography}
\end{document}